\documentclass{emulateapj}
\shorttitle{Star Formation and Extinction in $z\sim2$ Galaxies}
\shortauthors{Reddy et~al.}
\slugcomment{Received 2005 October 06; Accepted 2006 February 23}

\begin{document}
\newcommand{\ebmv}{E(B-V)}
\newcommand{\sfr}{{\rm M}_{\odot} ~ {\rm yr}^{-1}}
\newcommand{\zmk}{(z-K)_{\rm AB}}
\newcommand{\jmk}{J-K_{\rm s}}
\newcommand{\rmk}{{\cal R}-\ks}
\newcommand{\ugr}{U_{\rm n}G{\cal R}}
\newcommand{\rs}{{\cal R}}
\newcommand{\bzk}{BzK}
\newcommand{\kab}{K_{\rm AB}}
\newcommand{\ks}{K_{\rm s}}
\newcommand{\lya}{Lyman~$\alpha$}
\newcommand{\lyb}{Lyman~$\beta$}
\newcommand{\za}{$z_{\rm abs}$}
\newcommand{\ze}{$z_{\rm em}$}
\newcommand{\cmtwo}{cm$^{-2}$}
\newcommand{\nhi}{$N$(H$^0$)}
\newcommand{\degpoint}{\mbox{$^\circ\mskip-7.0mu.\,$}}
\newcommand{\kms}{\,km~s$^{-1}$}      
\newcommand{\minpoint}{\mbox{$'\mskip-4.7mu.\mskip0.8mu$}}
\newcommand{\peryr}{\mbox{$\>\rm yr^{-1}$}}
\newcommand{\secpoint}{\mbox{$''\mskip-7.6mu.\,$}}
\newcommand{\sqdeg}{\mbox{${\rm deg}^2$}}
\newcommand{\squig}{\sim\!\!}
\newcommand{\subsun}{\mbox{$_{\twelvesy\odot}$}}
\newcommand{\et}{{\rm et al.}~}

\def\ltsima{$\; \buildrel < \over \sim \;$}
\def\simlt{\lower.5ex\hbox{\ltsima}}
\def\gtsima{$\; \buildrel > \over \sim \;$}
\def\simgt{\lower.5ex\hbox{\gtsima}}
\def\arcs{$''~$}
\def\arcm{$'~$}
\def\erf{\mathop{\rm erf}}
\def\erfc{\mathop{\rm erfc}}
\title{STAR FORMATION AND EXTINCTION IN REDSHIFT $z\sim 2$ GALAXIES:
INFERENCES FROM {\it SPITZER} MIPS OBSERVATIONS\altaffilmark{1}}
\author{\sc Naveen A. Reddy\altaffilmark{2}, Charles C. Steidel\altaffilmark{2},
Dario Fadda\altaffilmark{3}, Lin Yan\altaffilmark{3}, Max Pettini\altaffilmark{4},
Alice E. Shapley\altaffilmark{5}, Dawn K. Erb\altaffilmark{6}, and Kurt L. Adelberger\altaffilmark{7}}

\altaffiltext{1}{Based, in part, on data obtained at the W.M. Keck
Observatory, which is operated as a scientific partnership among the
California Institute of Technology, the University of California, and
NASA, and was made possible by the generous financial support of the
W.M. Keck Foundation.  Also based in part on observations made with the 
{\it Spitzer} Space Telescope, which is operated by the Jet Propulsion
Laboratory, California Institute of Technology under a contract with
NASA.}
\altaffiltext{2}{California Institute of Technology, MS 105--24, Pasadena, CA 91125}
\altaffiltext{3}{{\it Spitzer} Science Center, California Institute of Technology, MS 220--6, Pasadena, CA 91126}
\altaffiltext{4}{Institute of Astronomy, Madingley Road, Cambridge CB3 OHA, UK}
\altaffiltext{5}{Department of Astrophysical Sciences, Peyton Hall-Ivy Lane, Princeton, NJ 08544}
\altaffiltext{6}{Harvard-Smithsonian Center for Astrophysics, 60 Garden Street, Cambridge, MA 02138}
\altaffiltext{7}{McKinsey \& Company, 1420 Fifth Avenue, Suite 3100, Seattle, WA 98101}

\begin{abstract}

Using very deep {\it Spitzer} MIPS $24$~$\mu$m observations, we
present an analysis of the bolometric luminosities and UV extinction
properties of more than 200 spectroscopically identified, optically
selected ($\ugr$) $z \sim 2$ galaxies in the GOODS-N field.  The large
spectroscopic sample of rest-UV selected galaxies is supplemented with
photometrically identified near-IR-selected (``BzK'' and ``DRG'')
galaxies and sub-mm sources at similar redshifts in the same field,
providing a representative collection of relatively massive ($M^{\ast}
> 10^{10}$~M$_{\odot}$) galaxies at high redshifts. We focus on the
redshift range $1.5<z<2.6$, for which the $24$~$\mu$m observations
provide a direct measurement of the strength of the mid-IR PAH
features in the galaxy spectra; the rest-frame $5-8.5$~$\mu$m
luminosities ($L_{\rm 5-8.5\mu m}$) are particularly tightly
constrained for the objects in our sample with precise spectroscopic
redshifts.  We demonstrate, using stacked X-ray observations and a
subset of galaxies with H$ \alpha$ measurements, that $L_{\rm 5-8.5\mu
m}$ provides a reliable estimate of $L_{\rm IR}$ for most star forming
galaxies at $z \sim 2$. We show that the range of $L_{\rm IR}$ in the
optical/near IR-selected samples considered extends from $\simeq
10^{10}$~L$_{\odot}$ to $>10^{12}$~L$_{\odot}$, with a mean $\langle
L_{\rm IR} \rangle \simeq 2\times 10^{11}$~L$_{\odot}$.  The LIRG
population at $z\sim 2$ is essentially the same population of galaxies
that are selected by their optical/near-IR colors.  Objects with LIRG
to ULIRG luminosities are present over the full range of stellar
masses in the samples, from $2 \times 10^{9}$~M$_{\odot}$ to
$5\times10^{11}$~M$_{\odot}$.  We use the MIPS $24$~$\mu$m
observations for an independent examination of dust extinction in high
redshift galaxies, and demonstrate that, as in the local universe, the
obscuration ($L_{\rm IR}\over L_{\rm 1600}$) is strongly dependent on
bolometric luminosity, and ranges in value from $<1$ to $\sim 1000 $
within the sample considered. However, the obscuration is $\sim 10$
times smaller at a given $L_{\rm bol}$ (or, equivalently, a similar
level of obscuration occurs at luminosities $\sim 10$ times larger) at
$z \sim2$ than at $z\sim 0$. We show that the values of $L_{\rm IR}$
and obscuration inferred from the UV spectral slope $\beta$ generally
agree well with the values inferred from $L_{\rm 5-8.5\mu m}$ for
$L_{\rm bol} < 10^{12}$~L$_{\odot}$.  As found previously by several
investigators, for ``ultraluminous'' objects with $L_{\rm bol} >
10^{12}$~L$_{\odot}$ it is common for UV-based estimates to
underpredict $L_{\rm IR}$ by a factor of $\sim 10-100$.  Using the
specific star formation rate of galaxies (SFR per unit stellar mass)
as a proxy for cold gas fraction, we find a wide range in the
evolutionary state of galaxies at $z\sim 2$, from galaxies that have
just begun to form stars to those which have already accumulated most
of their stellar mass and are about to become, or already are,
passively-evolving.

\end{abstract}


\section{Introduction}
\label{sec:intro}

The most direct method currently available for tracing the bolometric
luminosities of high redshift star-forming galaxies ($z\ga 2$) has
been from their submillimeter emission (e.g., \citealt{smail97};
\citealt{hughes98}; \citealt{barger98}).  Unfortunately, current
sensitivity limits of bolometers and submillimeter wave
interferometers allow for only the most luminous starburst galaxies to
be detected at high redshifts via their dust emission.  Further
compounding the problem is the coarse spatial resolution provided by
such instruments, making it difficult to distinguish the
counterpart(s) to the submillimeter emission for subsequent followup,
although the recently developed method of radio-detection has been a
breakthrough in alleviating this problem for most, but not all, bright
submillimeter galaxies (e.g., \citealt{chapman05}).  Regardless, the
dust properties of the vast majority of star-forming galaxies at high
redshift remained uninvestigated until recently.

The rest-frame far-infrared (FIR) wavelength region is still
inaccessible for the {\it typical} galaxy at redshifts $z\ga 1$, so we
must look to other portions of the spectrum to directly examine dust
properties.  Our understanding of the mid-IR properties of local and
high redshift galaxies advanced considerably with the launch of the
{\it Infrared Space Observatory} (ISO), which was sensitive enough to
detect the mid-IR emission of $10^{11}$~L$_{\odot}$ galaxies at $z\sim
1$ (e.g., \citealt{flores99, elbaz02, pozzi04, rowan04}).  These
observations revealed the almost ubiquitous presence of mid-IR dust
emission features in star forming galaxies in both the local and
$z\sim 1$ universe, and suggested the possibility of using the mid-IR
dust emission of galaxies as a tracer of bolometric luminosity
(\citealt{boselli98}; \citealt{adel00}; \citealt{dale00};
\citealt{helou00}; \citealt{forster03}).

These advances now continue with the highly successful {\it Spitzer}
Space Telescope, providing the same sensitivity as ISO in probing dust
emission from $10^{11}$~L$_{\odot}$ galaxies at $z\sim 2$.  The
progress made with {\it Spitzer} is particularly important for
studying galaxies at $z\sim 2$ because this epoch was until recently
largely uninvestigated, yet is believed to be the most active in terms
of star formation and the build up of stellar and black hole mass
(e.g., \citealt{dickinson03}; \citealt{rudnick03}; \citealt{madau96};
\citealt{lilly96}; \citealt{steidel99}; \citealt{shaver96};
\citealt{fan01}; \citealt{dimatteo03}; \citealt{giavalisco96}).  The
sensitivity afforded by the {\it Spitzer} MIPS instrument allows us to
examine the {\it typical} $L^*$ galaxy at $z\sim 2$, rather than a
limited slice of the most luminous population, a problem which, as
alluded to before, limits the usefulness of submillimeter
observations.

It is fortuitous that the rest-frame mid-IR spectral features observed
in local and $z\sim 1$ star-forming galaxies are redshifted into the
{\it Spitzer} IRS spectral and MIPS imaging passbands at $z\sim 2$.
The mid-IR spectral region from $3-15$~$\mu$m is rich with emission
lines believed to arise from the stochastic heating of small dust
grains by UV photons (see review by \citealt{genzel00}).  These
unidentified infrared bands (UIBs) are generally attributed to the
${\rm C=C}$ and ${\rm C-H}$ stretching and bending vibrational modes
of a class of carbonaceous molecules called polycyclic aromatic
hydrocarbons (PAHs; e.g., \citealt{puget89}; \citealt{tielens99}),
which we assume hereafter.  In the typical spectrum of a star-forming
galaxy, these PAH emission lines, along with various fine-structure
metal and HI recombination lines (e.g., \citealt{sturm00}), are
superposed on a mid-IR continuum thought to result from dust emission
from very small grains, or VSGs \citep{desert90}.  In star-forming
galaxies, the global PAH emission is mainly attributed to UV radiation
from OB stars and has been found to correlate with global star
formation rate (e.g., \citealt{forster04}; \citealt{forster03};
\citealt{roussel01}), although variations with ionizing intensity and
metallicity are also observed (e.g., \citealt{engelbracht05};
\citealt{hogg05}; \citealt{alonso04}; \citealt{helou01};
\citealt{normand95}).

Until now, the only way to estimate the bolometric luminosities of
most galaxies at $z\sim 2$ independent of extinction was via their
stacked X-ray and radio emission: unfortunately these data are not
sufficiently sensitive to detect individual $L^*$ galaxies at $z\sim
2$ (e.g., \citealt{reddy04}; \citealt{nandra02}; \citealt{brandt01}).
The {\it Spitzer} data considered in this paper are useful in
assessing the bolometric luminosities of galaxies on an individual
basis.  One is still limited because detailed mid-IR spectroscopy is
feasible only for the most luminous galaxies at $z\sim 2$ (e.g.,
\citealt{yan05}; \citealt{houck05}), but $L^{*}$ galaxies at $z\sim 2$
(with $L_{\rm bol}\sim 10^{11}$~L$_{\odot}$) can be detected in deep
$24$~$\mu$m images.  We employ MIPS $24$~$\mu$m data to study the
rest-frame mid-IR properties of optical and near-IR selected galaxies
at redshifts $1.5\la z\la 2.6$.  We describe the optical, near-IR,
X-ray, and mid-IR data in \S~\ref{sec:data} and \ref{sec:midir}.  Our
large sample of spectroscopic redshifts for optically-selected
galaxies allows us to very accurately constrain the rest-frame mid-IR
fluxes of $z\sim 2$ galaxies.  In \S~\ref{sec:photoz}, we describe our
method for estimating photometric redshifts for near-IR samples of
galaxies where spectroscopy is less feasible.  The procedure for
estimating infrared luminosities from MIPS data is outlined in
\S~\ref{sec:f24toir}.  We discuss the infrared luminosity
distributions of $24$~$\mu$m detected and undetected sources in
\S~\ref{sec:lirdist} and \ref{sec:f24stack}.  The dust attenuation
properties of optical and near-IR selected $z\sim 2$ galaxies and the
correlation of these properties with bolometric luminosity are
discussed in \S~\ref{sec:attenuation} and \S~\ref{sec:dustobs}.  The
stellar populations and composite rest-frame UV spectral properties of
faint $24$~$\mu$m galaxies are discussed in \S~\ref{sec:undet}.  In
\S~\ref{sec:masses} we examine in more detail the mid-IR properties of
massive galaxies at $z\sim 2$.  We conclude in \S~\ref{sec:disc} by
discussing the viability of optical and near-IR color criteria in
selecting LIRGs and ULIRGs at $z\sim 2$ and what the {\it Spitzer}
MIPS observations can reveal about the mass assembly of galaxies at
high redshift.  A flat $\Lambda$CDM cosmology is assumed with
$H_{0}=70$~km~s$^{-1}$~Mpc$^{-1}$ and $\Omega_{\Lambda}=0.7$.

\section{Sample Selection and Ancillary Data}
\label{sec:data}

\subsection{Optical and Near-IR Selection}

\begin{deluxetable*}{lcccccccc}[!tbp]
\tabletypesize{\footnotesize}
\tablewidth{0pc}
\tablecaption{Properties of the Samples}
\tablehead{
\colhead{} &
\colhead{} &
\colhead{} &
\colhead{} &
\colhead{$\rho$\tablenotemark{c}} &
\colhead{} &
\colhead{} &
\colhead{} &
\colhead{}
\\
\colhead{Sample} &
\colhead{Limits} &
\colhead{$N_{\rm c}$\tablenotemark{a}} &
\colhead{$N_{\rm x}$\tablenotemark{b}} &
\colhead{(arcmin$^{-2}$)} &
\colhead{$N_{\rm s}$\tablenotemark{d}} &
\colhead{$\langle z\rangle$\tablenotemark{e}} &
\colhead{$N_{\rm g}$\tablenotemark{f}} &
\colhead{$f_{\rm m}$\tablenotemark{g}}}
\startdata
\\
$\ugr$ (BX/BM/LBG) & ${\cal R}<25.5$ & 1571 & 23 & $10.2\pm 0.3$ & 313 & $2.25\pm0.33$ & 219 & 0.65 \\
$\bzk$/SF & $\ks<21.0$ & 221 & 32 & $3.1\pm 0.2$ & 53 & $2.09\pm0.34$ & 82 & 0.82 \\
$\bzk$/PE & $\ks<21.0$ & 17 & 4 & $0.24\pm 0.06$ & 0 & $1.70\pm0.20$ & 13 & 0.54 \\
DRG\tablenotemark{h} & $\ks<21.0$ & 73 & 19 & $1.0\pm 0.1$ & 5 & $2.48\pm0.35$ & 24 & 0.71\\
\enddata
\tablenotetext{a}{Number of candidates.}
\tablenotetext{b}{Number of directly-detected X-ray sources, including spectroscopically-confirmed galaxies.} 
\tablenotetext{c}{Surface density of candidates.  Errors are computed assuming Poisson statistics.}
\tablenotetext{d}{Number of spectroscopically confirmed objects with $z>1.5$.  Note that
we only obtained spectra for those $\bzk$ and DRG galaxies that satisfy the
$\ugr$ criteria.}
\tablenotetext{e}{Mean redshift of sample for $z>1.5$.  For the $\bzk$/PE
sample, this is the mean redshift of the {\it photometric} redshift distribution observed
for $\bzk$/PE galaxies (e.g., \citealt{daddi04b};\citealt{reddy05a}).  For
the DRGs, this is the mean redshift of the {\it spectroscopic} redshift distribution observed for DRGs with $z>1.5$ in four of the fields of the $z\sim 2$ optical survey \citep{reddy05a}.}
\tablenotetext{f}{Number of non-AGN galaxies (i.e., those with no direct X-ray detections) with spectroscopic redshifts $1.5<z<2.6$.  For the $\bzk$/SF sample, this number includes both spectroscopically-confirmed $\bzk$/SF galaxies (all
of which are in the $\ugr$ sample) and those with secure photometric redshifts.  For the $\bzk$/PE sample, this
includes all candidates without direct X-ray detections.  For the  
DRG sample, this number includes galaxies with photometric redshifts $1.5<z<2.6$.}
\tablenotetext{g}{Fraction of MIPS $24$~$\mu$m detections to $8$~$\mu$Jy ($3$~$\sigma$) among the $N_{\rm g}$ galaxies.}
\tablenotetext{h}{The DRG sample includes both star-forming galaxies and those with
little star formation.  As discussed in \citet{reddy05a}, those with spectra
(i.e., those DRGs which also satisfy the $\ugr$ criteria) are likely
to be currently forming stars.\\}
\label{tab:stats}
\end{deluxetable*}

\begin{figure}[hbt]
\plotone{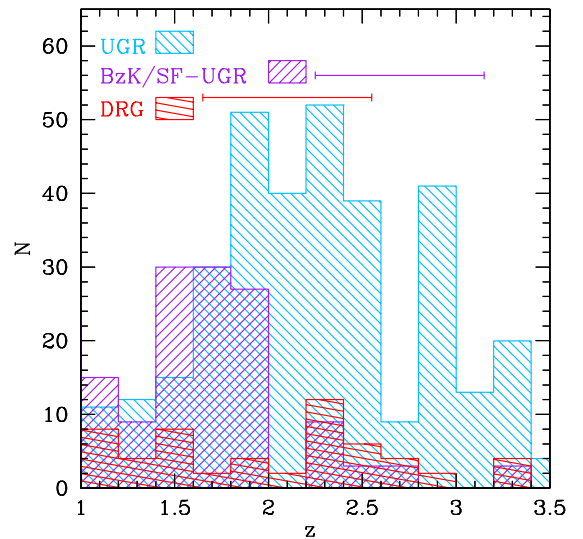}
\caption{{\it Spectroscopic} redshift distribution of optically (i.e.,
$\ugr$) selected galaxies, with typical error of $\sigma(z)=0.002$.
Also shown are arbitrarily normalized photometric redshift
distributions of $\bzk$/SF galaxies that do not satisfy the $\ugr$
criteria, and DRGs.  The errorbars indicate the average uncertainty in
photometric redshifts ($\sigma(z)=0.45$) for the near-IR selected
$\bzk$ galaxies and DRGs.\\
\label{fig:zhist}}
\end{figure}

The star-forming galaxies studied here were drawn from the sample of
$z\sim 2$ galaxies in the GOODS-N field selected based on their
observed $\ugr$ colors to ${\cal R}=25.5$ \citep{adel04,steidel04}.
The optical images used for the selection of candidates cover an area
$11\arcmin$ by $15\arcmin$.  We refer to ``BM'' and ``BX'' galaxies as
those which are selected to be at redshifts $1.5\la z\la 2$ and
$2.0\la z\la 2.6$, respectively (\citealt{adel04};
\citealt{steidel04}).  In addition to BX/BM galaxies, we also consider
galaxies selected using the $z\sim 3$ LBG criteria \citep{steidel03}.
The BX/BM and LBG candidates make up our $\ugr$, or optically,
selected sample.  We obtained rest-frame UV spectra with the blue
channel of the Low Resolution Imaging Spectrograph (LRIS-B) on Keck I
for 386 $\ugr$ candidates.  The numbers of candidates and
spectroscopically confirmed galaxies in the $\ugr$ sample are
summarized in Table~\ref{tab:stats}.  The {\it spectroscopic} redshift
distribution of $\ugr$ galaxies in the GOODS-N field is shown in
Figure~\ref{fig:zhist}.  For efficiency, we preferentially targeted
for spectroscopy those $\ugr$ candidates with ${\cal R}$-band
magnitudes in the range ${\cal R}=22.5-24.5$ (AB units) and gave lower
priorities for fainter objects where redshift identification is more
difficult from absorption lines and brighter objects where the
contamination fraction (from low redshift interlopers) is larger.  The
star formation rate distribution of spectroscopically-confirmed $\ugr$
galaxies is similar to that of all $\ugr$ galaxies in the targeted
redshift range to $\ks=21$ (Vega; \citealt{reddy05a}; see also
\citealt{steidel04} for a discussion of spectroscopic bias of galaxy
properties with respect to the photometric sample of $\ugr$ galaxies).

Our deep near-IR $J$ and $K$-band imaging, in addition to publicly
available deep optical imaging, allows us to select both star-forming
galaxies and those with little current star formation in the GOODS-N
field.  Details of the optical and near-IR images are provided in
\citet{reddy05a}.  The near-IR selection of star-forming galaxies is
done using the criteria of \citet{daddi04b}, resulting in a
``$\bzk$/SF'' sample (e.g., \citealt{daddi04b,reddy05a}).  The near-IR
selected samples of galaxies with very little current star formation
are constructed by considering the $\bzk$ and $\jmk$ colors of
candidates satisfying the $\bzk$/PE and Distant Red Galaxy (DRG)
criteria (e.g., \citealt{reddy05a}; \citealt{daddi04b};
\citealt{franx03}).  Approximately $70\%$ of DRGs to $\ks=21$ (Vega)
have signatures of intense star formation (\citealt{papovich05};
\citealt{reddy05a}; \citealt{vandokkum04})\footnote{In order to ensure
our sample is complete, we only consider DRGs to $\ks=21$.}.  The $J$
and $K$ band images cover a large area by near-IR standards ($\sim
8.5\arcmin\times 8.5\arcmin$), but are still less than half the area
of the optical $\ugr$ images.  The number of near-IR selected
candidates and their surface densities are presented in
Table~\ref{tab:stats}.  For the remaining analysis, we use AB units
for optical ($\ugr$) magnitudes and Vega units for near-IR ($\ks$)
magnitudes.

\subsection{X-Ray Data}

The very deep {\it Chandra} X-ray data in the GOODS-N field
\citep{alexander03} allow for an independent means of assessing the
presence of AGN in the samples, which can be quite significant for
near-IR selected samples \citep{reddy05a}.  In addition, we can stack
the X-ray data for those galaxies lying below the X-ray detection
threshold to determine their average emission properties
(e.g. \citealt{laird05}; \citealt{lehmer05}; \citealt{reddy04};
\citealt{nandra02}; \citealt{brandt01}).  The X-ray data and stacking
methods are discussed in detail in \citet{reddy04} and
\citet{reddy05a}.  The numbers of directly-detected X-ray sources in
each sample considered here are summarized in Table~\ref{tab:stats}.\\

\section{Mid-IR Data}
\label{sec:midir}

The mid-IR data are obtained from the {\it Spitzer} Multiband Imaging
Photometer (MIPS) instrument.  The $24$ micron data were taken as part
of the GOODS Legacy Survey (P.I.: M. Dickinson) between May 27 and
June 6, 2004. They consist of 24 AORs (Astronomical Observation
Requests) of approximately 3 hours each.  The combined data reach a
depth equivalent to $\sim 10$ hours integration at any point in the
mosaicked image.  The data are publicly available since August 2004 in
the Spitzer archive.  The basic calibrated data (BCD) produced by the
Spitzer pipeline were used as the starting point for the data
reduction.  As explained in detail by \citet{fadda06}, there were
several artifacts which added noise to the images, hampering the
detection of faint sources.  These artifacts include image latencies
from previous observations of bright objects or image dark spots
present on the pick-off mirror that are projected in different
positions by the cryogenic scan mirror during observations.  Other
variations come from the variable zodiacal light.  We have corrected
each BCD image for these effects using the procedure described by
\citet{fadda06}.  The final mosaic consists of 7198 BCDs combined
using MOPEX \citep{makovoz05}.  The final reduced $24$~$\mu$m mosaic
of the GOODS-N region has a pixel size of $1\farcs275$ and covers the
entirety of our optical $\ugr$ images and the measured $3$~$\sigma$
depth is approximately $8$~$\mu$Jy.  The large beam-size of MIPS
($\sim 5\farcs 4$ at $24$~$\mu$m) combined with the effects of
blending make aperture photometry impractical.  Instead, we have
chosen to use a PSF fitting method to extract $24$~$\mu$m fluxes for
our galaxies, similar to the method used to extract IRAC fluxes for
galaxies as discussed in \citet{shapley05}.

We first compute an empirical PSF using a two-pass approach.  In the
first pass, we take the median flux profile of several tens of
distinct MIPS point sources across the mosaic to create a rough
empirical PSF.  This PSF is then used to subtract sources in proximity
to the tens of MIPS point sources in order to better determine the PSF
profile out to larger radii.  This is important for the MIPS PSF which
contains up to $20\%$ of the point source flux in the Airy rings.  The
effects of source confusion are mitigated by employing the higher
spatial resolution {\it Spitzer} IRAC data in the GOODS-N field to
constrain the MIPS source positions.  The empirical PSF, normalized to
unit flux, is fit to these positions and the fluxes are extracted.
The PSF used here extends to $15\farcs 3$ radius, and we apply a
$15\%$ aperture correction based on the observed curves-of-growth of
MIPS point source profiles from the First Look Survey.

The number of MIPS detections ($>3$~$\sigma$) and non-detections in
each sample are summarized in Table~\ref{tab:stats}.  Virtually all of
the directly-detected X-ray sources are detected at $24$~$\mu$m and
all have optical/X-ray flux ratios indicating that AGN dominate the
X-flux \citep{reddy05a}.  Submillimeter galaxies are often associated
with direct X-ray detections (e.g., \citealt{alexander05};
\citealt{reddy05a}) even though their bolometric luminosities may
still be dominated by star-formation.  Since we are primarily
interested in the mid-IR emission as a tracer of star formation, we
have excluded all directly-detected X-ray sources (almost all of which
are AGN; \citealt{reddy05a}) for most of the analysis, unless they
happen to coincide with a radio-detected SMG from the
\citet{chapman05} (SC05) sample.  We caution that the resulting sample
of $9$ SMGs is not meant to be uniform or complete: $\sim 40\%$ of
SMGs are not associated with radio sources, either because of their
higher redshifts or radio faintness (e.g., \citealt{chapman05};
\citealt{smail02}).  Furthermore, the submillimeter observations are
not uniform over the field.  Nonetheless, it is of obvious interest to
at least compare the mid-IR properties of this limited set of
radio-detected SMGs to those of galaxies in other samples.

\begin{figure*}[hbt]
\plotone{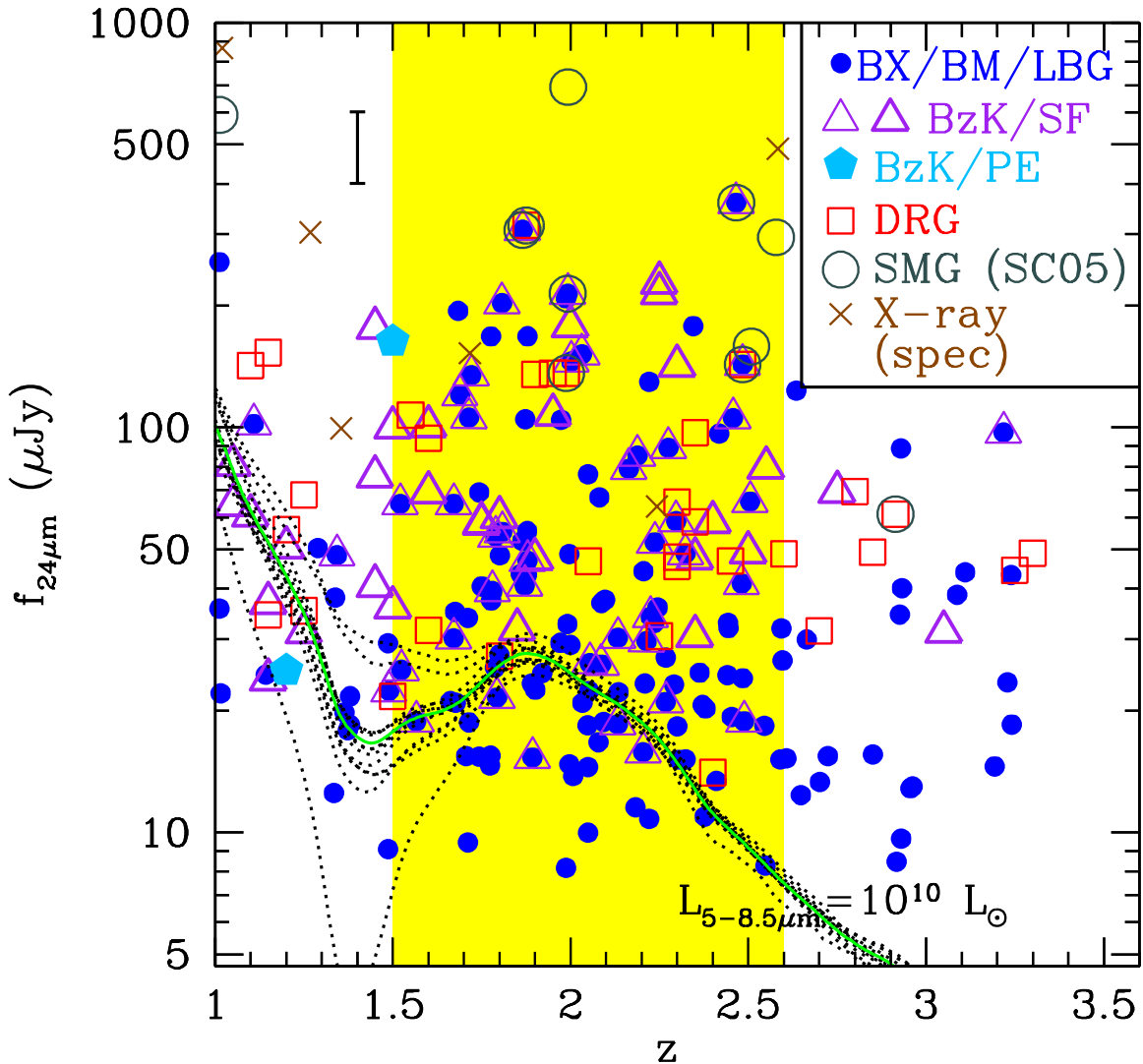}
\caption{Expected $24$~$\mu$m fluxes of the local and high redshift
  template galaxies summarized in Table~\ref{tab:templates} shown by
  the dotted curves, normalized so that $L_{\rm 5-8.5\mu
  m}=10^{10}$~L$_{\odot}$.  The solid curve is the average over all
  the templates.  Also shown are the observed fluxes of $\ugr$ (small
  points) and $\bzk$/SF galaxies (thin triangles) with spectroscopic
  redshifts; and $\bzk$/SF (thick triangles), $\bzk$/PE (solid
  pentagons), and DRGs (open squares) with photometric redshifts.  The
  typical error in photometric redshifts of DRG and $\bzk$ galaxies is
  $\sigma(z)\sim 0.45$.  Radio-detected submillimeter galaxies to
  $S_{\rm 850\mu m}\sim 4.2$~mJy from \citet{chapman05} are shown by
  the open circles.  We have removed directly-detected X-ray sources
  in the samples above, except for those in the SMG sample.  Crosses
  denote hard-band X-ray sources with spectroscopic redshifts in the
  $\ugr$ sample.  The vertical errorbar in the upper left-hand side of
  the figure shows the typical uncertainty in the $24$~$\mu$m flux.
  The shaded region indicates the redshift range over which the MIPS
  $24$~$\mu$m filter directly samples the rest-frame $7.7$~$\mu$m PAH
  feature.\\
\label{fig:pahvf24}}
\end{figure*}

The MIPS $24$~$\mu$m filter directly samples the rest-frame luminosity
from $5-8.5$~$\mu$m ($L_{\rm 5-8.5\mu m}$) for redshifts $1.5\la z\la
2.6$.  We used the mid-IR spectral shapes of star-forming galaxies
(listed in Table~\ref{tab:templates}) as templates in order to {\it
K}-correct the $24$~$\mu$m fluxes to determine $L_{\rm 5-8.5\mu m}$.
Figure~\ref{fig:pahvf24} shows the expected $24$~$\mu$m fluxes of the
galaxies listed in Table~\ref{tab:templates} as a function of
redshift.  These galaxies were chosen to cover a large range in SFRs
(from quiescent spiral galaxies to starbursts and LIRGs/ULIRGs).  The
mid-IR spectra are obtained from either ISO (for local galaxies;
\citealt{forster04b}) or {\it Spitzer} IRS (for $z\sim 2$
hyper-luminous galaxies; \citealt{yan05}) observations.  Some
properties of these galaxies are summarized in
Table~\ref{tab:templates}.  The mid-IR spectrum of each galaxy is
redshifted, convolved with the MIPS $24$~$\mu$m filter, and the fluxes
are normalized to have $L_{\rm 5-8.5\mu m}= 10^{10}$~L$_{\odot}$ to
produce the dotted curves in Figure~\ref{fig:pahvf24}.  The small
dispersion between the templates over redshifts $1.5\la z\la 2.6$
reflects small changes in the broadband $24$~$\mu$m fluxes of galaxies
due to changes in the relative strengths of the various PAH emission
lines and the ratio of PAH-to-continuum flux.  The solid curve in the
figure shows the average of all the template galaxies.  The prominent
peak at $z\approx 1.9$ is primarily due to the $7.7$~$\mu$m feature
shifting into the MIPS $24$~$\mu$m filter.\\

\begin{deluxetable}{lcc}[!tbp]
\tabletypesize{\footnotesize}
\tablewidth{0pc}
\tablecaption{Local Template Galaxies}
\tablehead{
\colhead{} &
\colhead{} &
\colhead{$L_{\rm IR}$} \\
\colhead{Name\tablenotemark{a}} &
\colhead{$z$\tablenotemark{b}} &
\colhead{($\times 10^{10}$~L$_{\odot}$)}}
\startdata
\\
M82 & 0.000677 & 4.8 \\
NGC253 & 0.000811 & 1.8 \\
M83 & 0.001711 & 1.9 \\
M51 & 0.002000 & 2.4 \\
NGC1808 & 0.003319 & 3.8 \\
NGC1097 & 0.004240 & 3.8 \\
NGC1365 & 0.005457 & 8.7 \\
NGC520 & 0.007609 & 6.5 \\
NGC7714 & 0.009333 & 5.6 \\
NGC3256 & 0.009386 & 40 \\
NGC6240 & 0.024480 & 60 \\
IRS9 & 1.83 & 1800 \\
\enddata
\label{tab:templates}
\tablenotetext{a}{All galaxies are compiled from \citet{forster04b}
and \citet{roussel01}, except for IRS9 and NGC7714 which are taken 
from \citet{yan05}.  The bolometric luminosity of IRS9 is constrained within
a factor of 2-3 \citep{yan05}.}
\tablenotetext{b}{Redshifts for local galaxies are obtained from
NASA/IPAC Extragalactic Database (NED).\\}
\end{deluxetable}

\section{Photometric Redshifts of Near-IR Selected Galaxies}
\label{sec:photoz}

Figure~\ref{fig:pahvf24} illustrates the sensitive dependence of the
{\it K}-correction on the redshift (e.g., galaxies with a given
observed $f_{\rm 24\mu m}$ can have a factor of 3 spread in $L_{\rm
5-8.5\mu m}$ depending on their redshift in the range $1.5<z<2.6$).
Our large {\it spectroscopic} sample gives us the distinct advantage
of knowing the {\it precise} redshifts for the optically selected
galaxies, removing the added uncertainty introduced by photometric
redshifts where the precise location of the PAH features with respect
to the MIPS $24$~$\mu$m filter is unknown, adding considerable
uncertainty to the inferred infrared luminosities (e.g.,
\citealt{papovich05}).  As we show below, the typical error in the
photometric redshifts derived for near-IR selected galaxies (even when
using data across a large baseline in wavelength, from UV through {\it
Spitzer} IRAC), is $\sigma(z)\sim 0.5$.  This error in redshift
translates to at least a factor of three uncertainty in $L_{\rm
5-8.5\mu~m}$.

Nonetheless, photometric redshifts are the only practical option for
optically faint galaxies where spectroscopy is not feasible.  This is
true for many of the DRGs and $\bzk$/PE galaxies.  We supplement our
spectroscopic database of optically selected galaxies with photometric
redshifts of near-IR selected galaxies.  We made use of the HyperZ
code to determine photometric redshifts \citep{bolzonella00}.  To
quantify the uncertainties in photometric redshifts, we tested the
code on $\ugr$ galaxies with spectroscopic redshifts, fitting to the
$U_{\rm n}BGV{\cal R}Iz+JK$ photometry.  The $BVIz$ magnitudes are
obtained from the v1.1 release of the GOODS ACS catalogs
\citep{giavalisco04}.  Errors in the optical $\ugr$ and near-IR $JK$
magnitudes are determined from Monte Carlo simulations described by
\citet{erb06b} and \citet{shapley05}.

\begin{figure*}[!tbh]
\plottwo{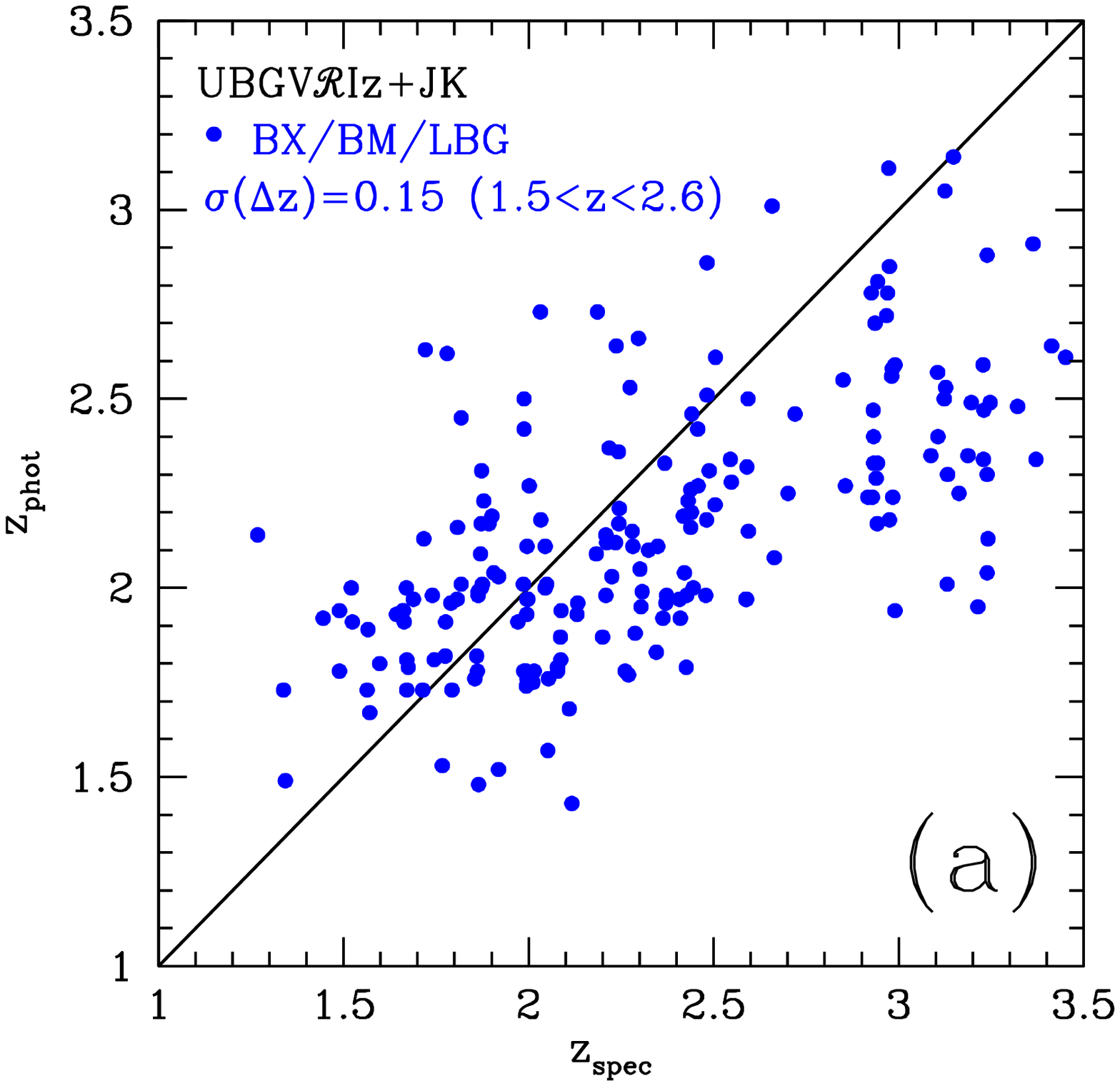}{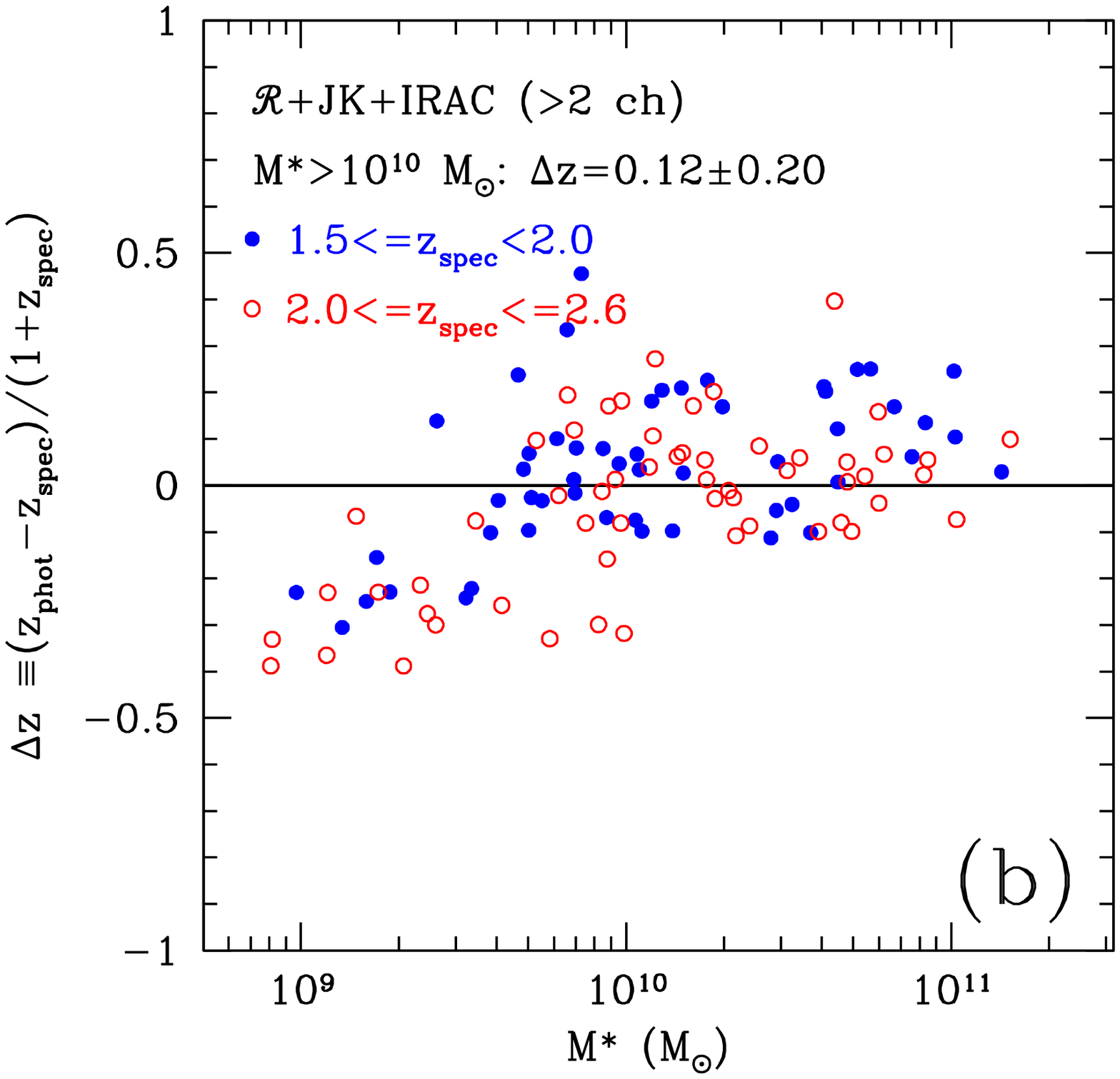}
\caption{{\it Left panel:} Photometric versus spectroscopic redshifts
for $\ugr$ galaxies. The solid line indicates $z_{\rm phot} = z_{\rm
spec}$.  {\it Right panel:} $\Delta z$, as defined in the text, as a
function of stellar mass for galaxies with redshifts $1.5\le z<2.0$
(solid circles) and $2.0\le z\le 2.6$ (open circles).  The figure only
shows objects with $>3$~$\sigma$ detections in at least two IRAC
channels.\\
\label{fig:zphot}}
\end{figure*}

The $\chi^2$ between the modeled and observed colors was calculated
for each galaxy for a number of different star formation histories
(with exponential decay times $\tau=10$, $20$, $50$, $100$, $200$,
$500$, $1000$, $2000$, and $5000$~Myr, and $\tau=\infty$) as a
function of redshift.  The best-fit photometric redshift is the
redshift at which $\chi^2$ is minimized.  As a figure of merit of the
resulting fit (and for easy comparison with other studies), we compute
\begin{eqnarray}
\Delta z \equiv (z_{\rm phot}-z_{\rm spec})/(1+z_{\rm spec}).
\end{eqnarray}
The results are shown in the left panel of Figure~\ref{fig:zphot}.
The dispersion in $\Delta z$ is $\sigma(\Delta z)\approx 0.15$ for
galaxies with spectroscopic redshifts $1.5<z<2.6$.  The actual error
in redshift is typically $\sigma(z)\sim 0.45$.  Both the
\citet{shapley05} code (which uses \citet{bruzual03} models) and the
\citet{benitez00} code gave similar results.  We found that adding
{\it Spitzer} IRAC data does little to tighten the scatter between
photometric and spectroscopic redshifts for most galaxies, reflecting
the absence of any distinct features (e.g., strong spectral breaks)
across the IRAC bands for $z\sim 2$ galaxies.  We note from the left
panel of Figure~\ref{fig:zphot} that photometric redshifts
systematically underestimate the true redshifts of $z>3$ galaxies.
This should not significantly affect our subsequent analysis since we
only consider galaxies up to $z=2.6$, and most of the $\bzk$/SF and
DRG galaxies have photometric redshifts $z_{\rm phot} \la 2.5$.

The IRAC data are nonetheless a powerful tool in discerning the more
massive galaxies from the less massive ones.  Since most of the
optically-faint DRGs and $\bzk$/PE galaxies are on average amongst the
more massive galaxies at $z\sim 2$ (e.g., Figure~18 of
\citet{reddy05a}), we have incorporated the IRAC data in the photometric
redshift fits.  The right panel of Figure~\ref{fig:zphot} shows
$\Delta z$ as a function of inferred stellar mass for $\ugr$ galaxies
computed for the best-fit $\tau$ model, normalizing to the observed
colors\footnote{We assume a Salpeter IMF in calculating the stellar
mass.}.  The scatter in $\Delta z$ for galaxies with stellar masses
$M^{\ast} \ga 10^{10}$~M$_{\odot}$ is $\sigma(\Delta z)\sim 0.20$, and we
will assume this value for the error in photometric redshifts of the
DRG and $\bzk$/PE galaxies.  To extend the comparison presented by
\citet{reddy05a} between $\ugr$ and $\bzk$ selected samples of
star-forming galaxies by examining their mid-IR properties, we compute
photometric redshifts for $\bzk$/SF galaxies that do not satisfy the
$\ugr$ criteria.  For the $\bzk$/SF galaxies, we assume an error of
$\sigma(\Delta z)\sim 0.15$, according to Figure~\ref{fig:zphot}a.

We obtained 51 secure photometric redshift fits for $\bzk$/SF galaxies
not selected by the $\ugr$ criteria (out of 95 such objects), and
their (arbitrarily normalized) photometric redshift distribution is
shown in Figure~\ref{fig:zhist}.  Also shown is the (arbitrarily
normalized) photometric redshift distribution for 28 (out of 49) non
X-ray detected DRGs for which we were able to derive secure
photometric redshifts $1<z_{\rm phot}<3.5$.  The remaining DRGs either
have $z_{\rm phot}<1$ or $z_{\rm phot}>3.5$ (and are irrelevant to the
analysis considered below) or had photometry that was inconsistent
with the \citet{bruzual03} models considered here, resulting in a
large $\chi^2$ value between the model and observed colors.  The DRGs
examined here appear to span a very large range in redshift from
$z\sim 1-3.5$, a result similar to that found by \citet{papovich05}
for DRGs in the GOODS-South field.  We obtained good photometric
redshift fits for only two of the $\bzk$/PE galaxies: one at $z\sim
1.2$ and the other at $z\sim 1.5$.  We reiterate that for purposes of
redshift identification, we only assumed the photometric redshifts for
those galaxies for which we were able to obtain good SED fits (i.e.,
with $\chi^2 \approx 1$) to the observed data.  There were a number of
objects for which the photometric redshift errors were relatively
large ($\delta z/(1+z)\ga 1$) or had derived redshifts that were much
larger ($z>4$) or smaller ($z<1$) than of interest here, and we
excluded such objects from our analysis.  Hence, the photometric
redshift distributions in Figure~\ref{fig:zhist} should not be
attributed to the populations as a whole.  For the remaining analysis
we consider only optically-selected galaxies with spectroscopic
redshifts and near-IR selected galaxies with photometric redshifts
between $1.5\la z\la2.6$ where the $24$~$\mu$m fluxes directly trace
the flux at rest-frame $7.7$~$\mu$m.  This is indicated by the shaded
region in Figure~\ref{fig:pahvf24}.

\section{Infrared Luminosities of Optical, Near-IR, and Submillimeter
Selected Galaxies at $z\sim 2$}

\subsection{Inferring Infrared Luminosities from $L_{\rm 5-8.5\mu m}$}
\label{sec:f24toir}

\begin{figure}[hbt]
\plotone{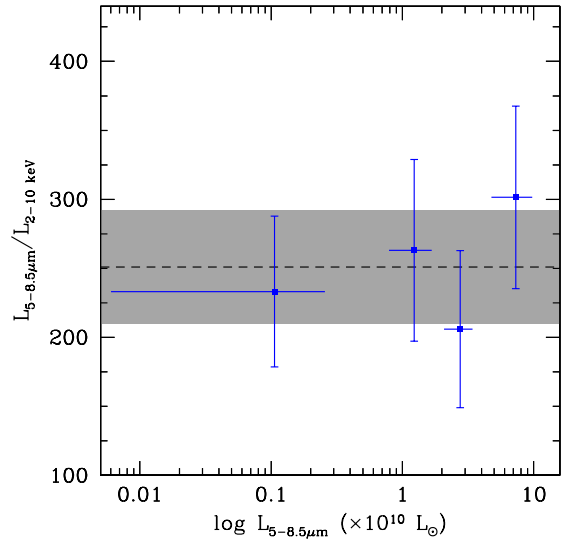}
\caption{Ratio of $L_{\rm 5-8.5\mu m}$ and stacked rest-frame
$2-10$~keV X-ray luminosity in bins of $L_{\rm 5-8.5\mu m}$ for
galaxies with redshifts $1.5<z<2.6$ and those with inferred ages
greater than $100$~Myr.  We have excluded sources directly-detected in
the {\it Chandra} 2~Ms data.  The faintest bin is for galaxies
undetected at $24$~$\mu$m.  Horizontal error bars indicate the
$1$~$\sigma$ dispersion in $L_{\rm 5-8.5\mu m}$ in each bin.  Vertical
error bars show the uncertainty in the mid-IR-to-X-ray luminosity
ratio, computed as the uncertainty in the mean X-ray luminosity added
in quadrature with the Poisson error in the mean mid-IR luminosity of
galaxies in each bin.  The dashed horizontal line and shaded region
indicate the mean and $1$~$\sigma$ uncertainty of the mid-IR-to-X-ray
luminosity ratio of $\sim 251 \pm 41$.\\
\label{fig:pahvxray}}
\end{figure}

The conversion between $L_{\rm 5-8.5\mu m}$ and infrared luminosity
will largely depend on the assumed spectral template relating the
mid-IR emission of galaxies to their total infrared luminosities.
Fortunately, the deep X-ray data allow us to determine whether $L_{\rm
5-8.5\mu m}$ scales with infrared luminosity (or star formation rate)
independent of any assumed template, adopting the local empirical
relationship between X-ray and FIR luminosity for star-forming
galaxies.  Figure~\ref{fig:pahvxray} shows the ratio of $L_{\rm
5-8.5\mu m}$ to stacked X-ray luminosity of (X-ray undetected)
galaxies in bins of $L_{\rm 5-8.5\mu m}$: we only considered
optically-selected galaxies with spectroscopic redshifts since it is
for these galaxies which we are able to most accurately constrain the
rest-frame X-ray luminosities.  Since X-ray emission is sensitive to
star formation on time-scales of $\ga 100$~Myr (see
\S~\ref{sec:attenuation}), Figure~\ref{fig:pahvxray} shows results
excluding galaxies with inferred ages $<100$~Myr.  Each bin contains
$10-20$ sources except the faintest bin which includes 45 galaxies
undetected at $24$~$\mu$m with ages $>100$~Myr.  The X-ray data for
galaxies in each bin were stacked using the procedure described in
\citet{reddy05a}.  The mean value of the mid-IR-to-X-ray luminosity
ratio is $\langle L_{\rm 5-8.5\mu m}/L_{\rm 2-10~keV}\rangle \approx
251\pm 41$.  The X-ray luminosities of local star-forming galaxies are
found to tightly correlate with their infrared emission for galaxies
with $10^{8}\la L_{\rm FIR}\la 10^{12}$~L$_{\odot}$ (e.g.,
\citealt{ranalli03}).  Using the X-ray luminosity as a proxy for
infrared luminosity therefore implies that the rest-frame mid-IR
fluxes follow the total infrared luminosity ($L_{\rm IR}$) for the
vast majority of optically-selected galaxies at $z\sim
2$\footnote{Another commonly used definition of $L_{\rm IR}$ is the
total luminosity from $1-1000$~$\mu$m.  This will differ from $L_{\rm
8-1000\mu m}$ by only a few percent, and for the remaining analysis,
we take $L_{\rm IR} \equiv L_{\rm 8-1000\mu m}$, as defined by
\citet{sanders96}.}.  The mid-IR fluxes must also follow the infrared
luminosity for most near-IR selected star-forming galaxies as well
given the large overlap ($70\%-80\%$) between optical and near-IR
selected samples of $z\sim 2$ star-forming galaxies \citep{reddy05a}.
As we show below, the conversion we assume between rest-frame mid-IR
and infrared luminosities reproduces the average infrared luminosities
predicted from stacked X-ray analyses (\S~\ref{sec:optical}).

To quantify the relationship between rest-frame mid-IR and total
infrared luminosity with a minimum number of assumptions, we have made
use of the data compiled by \citet{elbaz02}, which include IRAS $60$
and $100$~$\mu$m measurements and {\it ISO} observations of 149 local
star-forming galaxies with $L_{\rm IR}$ in the range $10^{9}\la L_{\rm
IR}\la 10^{12.6}$~L$_{\odot}$.  The mean and dispersion of the IR/MIR
flux ratio for the sample of 149 galaxies is $\langle \log (L_{\rm
IR}/L_{\rm 5-8.5\mu m})\rangle = 1.24\pm 0.35$.  Note the large
$1$~$\sigma$ dispersion of a factor of 2.2 in the IR/MIR flux ratio.
The dispersion in the IR/MIR flux ratios between galaxies may be
driven partly by changes in the mid-IR line strengths as the aromatic
carriers are dehydrogenated and/or destroyed depending on the
intensity of the ambient UV ionizing field (e.g., \citealt{alonso04};
\citealt{helou01}; \citealt{dale01}; \citealt{normand95}).
Metallicity effects and a changing distribution of dust with respect
to HII regions also likely contribute to the observed factor of $2-3$
dispersion in the IR/MIR ratios.  Nonetheless, the mean IR/MIR flux
ratio is similar to that inferred from the \citet{dale01} template SED
for a median $\log[f_\nu(60\mu m)/f_\nu(100\mu m)] \sim -0.20$.  Based
on the sample of 149 galaxies from \citet{elbaz02}, we assume $L_{\rm
IR}\approx 17.2 L_{\rm 5-8.5\mu m}$ to convert $L_{\rm 5-8.5\mu m}$ to
$L_{\rm IR}$.

It is worth noting that the relationship between $L_{\rm 5-8.5\mu m}$
and total IR luminosity for {\it local} star-forming galaxies may be
described by a more complicated function, such as a two power-law fit
(e.g., \citealt{elbaz02}).  These complicated relationships may not
apply to galaxies at $z\sim 2$ for several reasons.  First, the IR/MIR
ratio may change depending on the relative contribution of older
stellar populations in heating PAH and larger dust grains.  The
heating of dust by cooler stars is expected to be more prevalent for
the less luminous local galaxies with older stellar populations, on
average, than for $z\sim 2$ galaxies with relatively younger stellar
populations.  Second, it is found that $z\sim 2$ galaxies have
metallicities that are on average $0.3$~dex lower than those of local
galaxies at a given stellar mass \citep{erb06a}.  Therefore, the
metallicity dependence of the IR/MIR ratio found for local galaxies
(e.g., \citealt{engelbracht05}) suggests that we may not be able to
ascribe the IR/MIR ratio for a galaxy of a given stellar mass at $z=0$
to a galaxy at $z\sim 2$ with the same stellar mass.  A third
possibility, and one that is suggested by the results of this paper
and other studies (e.g., \citealt{adel00,calzetti99}), is that the
dust obscuration of galaxies at a given bolometric luminosity changes
as function of redshift, a result that may reflect dust enrichment
and/or a changing configuration of dust as galaxies age.  Therefore,
the relative distribution of PAH and larger dust grains within
galaxies may also change as a function of redshift.  Because of these
uncertainties, and since the primary motivation of our study is to
{\it independently} establish the validity of MIPS observations to
infer the infrared luminosities of $z\sim 2$ galaxies, we adopted the
simplest conversion that assumes only that the typical IR/MIR
luminosity ratio for local galaxies with a wide range in infrared
luminosity applies at $z\sim 2$.  By taking an average over the local
galaxies, we ensure that the derived $L_{\rm IR}$ is not more than a
factor of $2-3$ away from that predicted using the IR/MIR ratio of any
individual galaxy.  As we show below, our constant conversion
reproduces within the uncertainties the results that we obtain from
stacked X-ray and dust-corrected UV estimates of $L_{\rm IR}$.
 
\begin{figure}[hbt]
\plotone{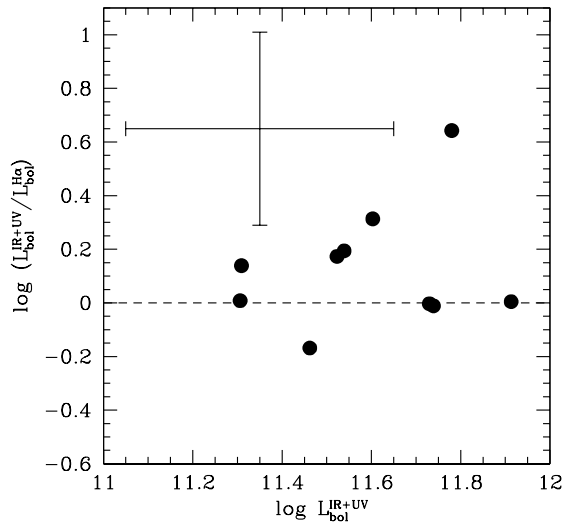}
\caption{Comparison between MIPS-inferred $L_{\rm bol}^{\rm IR+UV}$
and H$\alpha$-inferred $L_{\rm bol}^{\rm H\alpha}$ for a sample of
$10$ $\ugr$ galaxies.  The dispersion in the ratio of the two
estimates is $\sim 0.2$~dex for the subsample considered here.  The
errorbar shows the typical uncertainty of each point.\\
\label{fig:lbolcomp}}
\end{figure}

In addition to the stacked X-ray and dust-corrected UV estimates, we
also have spectroscopic H$\alpha$ observations for a small sample of
10 $\ugr$ galaxies in the GOODS-North field \citep{erb06c} with clean
(i.e., not blended) MIPS detections.  Once corrected for extinction,
the H$\alpha$ fluxes of these galaxies provide estimates of their
total (bolometric) luminosities, which we take to be the sum of the
$L_{\rm IR}$ and observed $1600$~\AA\, luminosity (uncorrected for
extinction):
\begin{eqnarray}
L_{\rm bol} \equiv L_{\rm IR} + L_{\rm 1600}.
\label{eq:bol}
\end{eqnarray}
In Figure~\ref{fig:lbolcomp} we show the comparison between $L_{\rm
bol}$ estimated from the sum of the MIPS-inferred $L_{\rm IR}$ and
observed $1600$~\AA\, luminosity ($L_{\rm bol}^{\rm IR+UV}$) with
$L_{\rm bol}$ estimated from the spectroscopic H$\alpha$ observations
($L_{\rm bol}^{\rm H\alpha}$).  The results indicate that within the
uncertainties the two estimates of $L_{\rm bol}$ track each other very
well (with a scatter of $0.2$~dex) over the range of $L_{\rm bol}$
typical of galaxies in optical/near-IR selected samples
(\S~\ref{sec:dustobs}), at least for this limited sample of $10$
objects.  The agreement between the MIPS and H$\alpha$-inferred
estimates suggests that our conversion relation between $L_{\rm
5-8.5\mu m}$ and $L_{\rm IR}$ works reasonably well.  Nonetheless, for
comparison with our constant conversion relation, we also consider the
effect on our results of assuming a two power-law conversion suggested
by \citet{elbaz02}.  As we show below, assuming the two power-law
conversion does not change the main conclusions of our study.

The far-infrared luminosity ($L_{\rm FIR}$) is typically defined to be
the luminosity from $40-120$~$\mu$m \citep{helou88}.  \citet{soifer87}
found $L_{\rm IR} \sim (1.91\pm0.17)\times L_{\rm FIR}$ for galaxies
in their Bright Galaxy Sample.  Modeling of the warm and cool
components of the dust emission in UV-bright galaxies indicates a
conversion factor of $\sim 1.75$ \citep{calzetti00}.  We take a median
value of $\sim 1.80$ in converting the inferred $L_{\rm IR}$ of
galaxies to a FIR luminosity.  Generally, uncertainties in the
conversion between $L_{\rm IR}$ and $L_{\rm FIR}$ are much smaller
than the uncertainties in converting $L_{\rm 5-8.5\mu m}$ to $L_{\rm
  IR}$.

Hereafter we assume uncertainties in the total infrared luminosities
as follows.  For $\ugr$ galaxies and radio-selected SMGs with
spectroscopic redshifts, we assume an uncertainty in $\log L_{\rm IR}$
of $0.3$~dex, corresponding to the dispersion in the conversion
between $L_{\rm 5-8.5\mu~m}$ and $L_{\rm IR}$.  For near-IR selected
$\bzk$ galaxies and DRGs, the photometric redshift error will add an
additional $0.5$~dex scatter.  The total uncertainty in $\log L_{\rm
  IR}$ for $\bzk$ galaxies and DRGs with photometric redshifts is
$0.6$~dex.

\subsection{Infrared Luminosity Distributions}
\label{sec:lirdist}

Figure~\ref{fig:pahvf24} summarizes the observed $f_{24\mu m}$ fluxes
of galaxies as a function of redshift.  In this figure, all direct
X-ray detections were removed from the $\ugr$, $\bzk$ and DRG samples,
unless they happened to coincide with a radio-detected submillimeter
galaxy (SMG) from \citet{chapman05}, or unless they have spectroscopic
redshifts in the $\ugr$ sample (crosses in Figure~\ref{fig:pahvf24}).
The $\ugr$ and SMG samples include objects outside the region covered
by our near-IR imaging.  The overlap between the samples considered
here is discussed extensively in \citet{reddy05a}.  The observed
$24$~$\mu$m fluxes for objects in the various samples generally span a
large range, from our sensitivity limit of $\sim 8$~$\mu$Jy to $\ga
200$~$\mu$Jy.

\begin{figure*}[hbt]
\plottwo{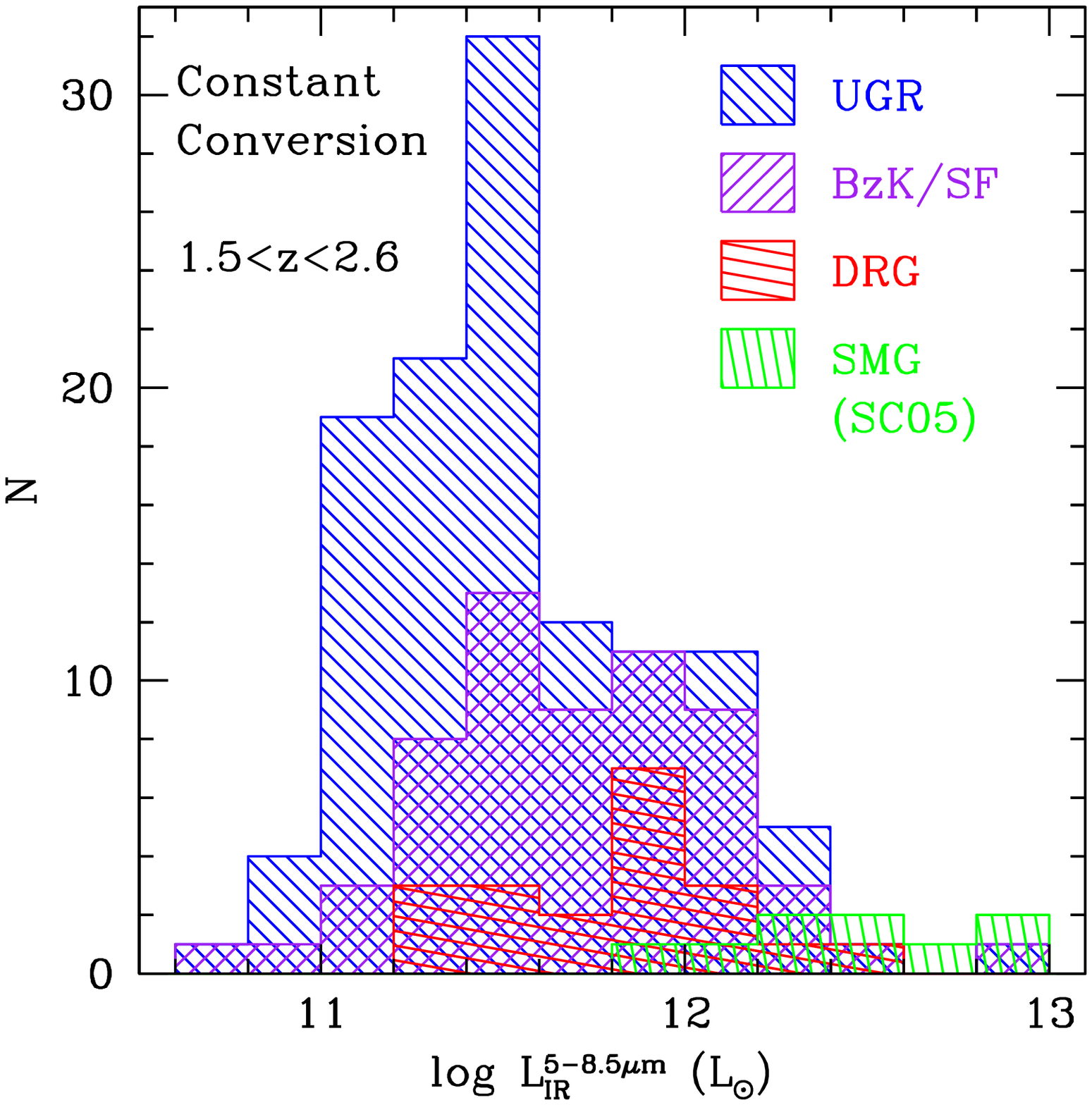}{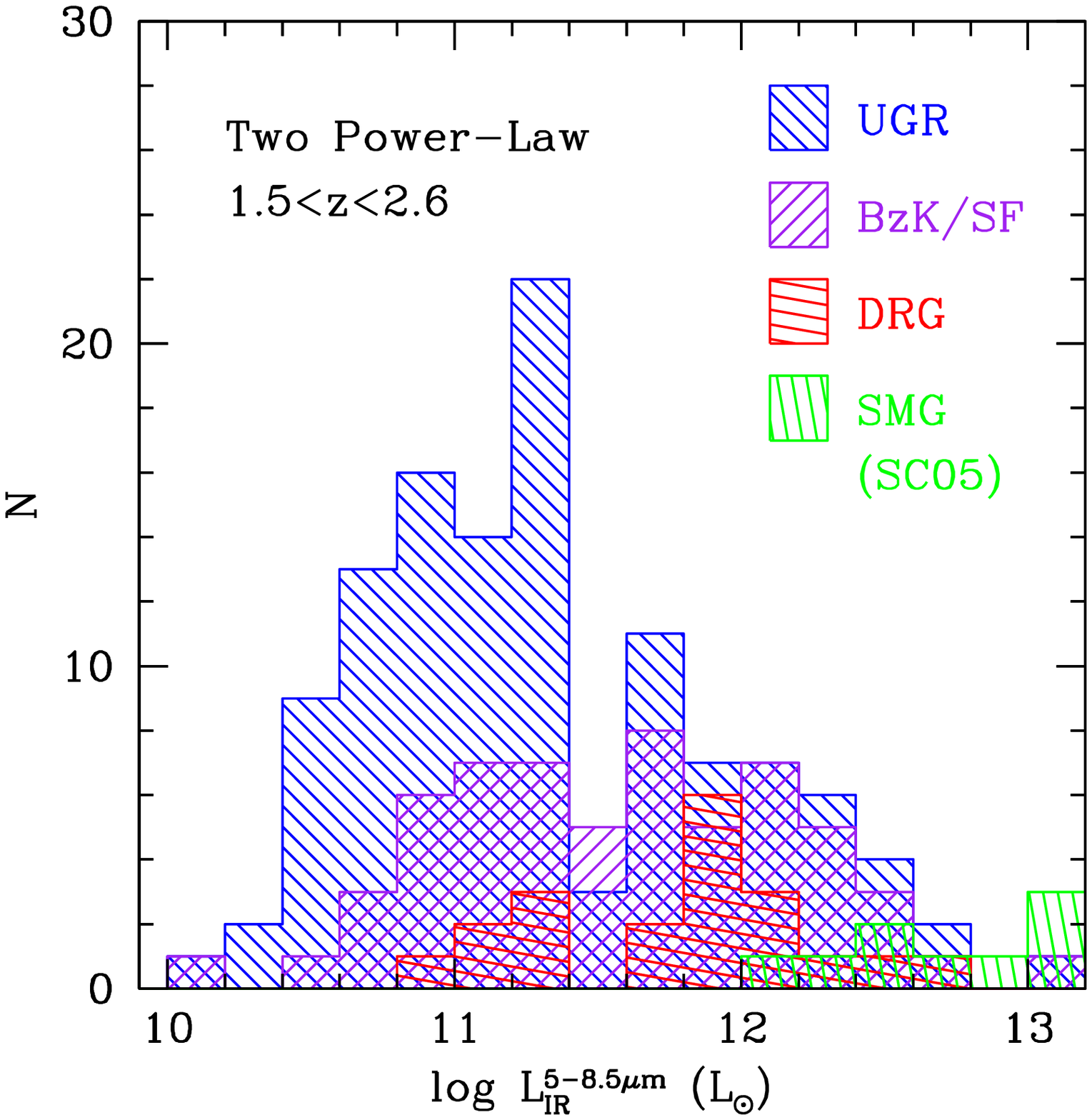}
\caption{Distributions of $L_{\rm IR}$ as inferred from $L_{5-8.5\mu
m}$ for galaxies in the $\ugr$, $\bzk$/SF, DRG, and radio-detected SMG
\citep{chapman05} samples with redshifts $1.5<z<2.6$ assuming our
constant conversion between $L_{\rm 5-8.5\mu m}$ and $L_{\rm IR}$
({\it left panel}) and the two power-law conversion of \citet{elbaz02}
({\it right panel}).  We have excluded directly-detected X-ray
sources unless they happen to coincide with a radio-detected SMG.\\
\label{fig:lirhist}}
\end{figure*}

For a more meaningful comparison, we have computed $L_{\rm IR}$ for
galaxies using the prescription described in \S~\ref{sec:f24toir}.
Figure~\ref{fig:lirhist} shows the distribution of $L_{\rm IR}$ as
inferred from $L_{\rm 5-8.5\mu m}$ for galaxies in the $\ugr$,
$\bzk$/SF, DRG, and radio-detected SMG \citep{chapman05} samples with
either spectroscopic or photometric redshifts $1.5<z<2.6$.  As in
Figure~\ref{fig:pahvf24}, we have excluded directly detected X-ray
sources from the distributions unless they coincide with a
radio-detected SMG source.  We show the resulting distributions
assuming the constant conversion and two power-law conversion
relations in the left and right panels, respectively.  The
distributions assuming the two power-law conversion are bimodal.
However, the distributions of observed UV luminosities and dust
correction factors of $\ugr$ galaxies are approximately gaussian
(e.g., \citealt{steidel04, reddy06}).  Assuming the \citet{calzetti00}
law to convert the observed luminosities to dust-corrected
luminosities then implies that the bolometric luminosity distribution
of $\ugr$ galaxies should be gaussian, a result not in accordance with
the bimodal distribution computed assuming the two power-law
conversion.  More generally, we expect to find gaussian distributions
of luminosity for galaxies in photometric surveys since such galaxies
are typically selected on a continuous range of observed colors and/or
magnitudes.  The bimodality in Figure~\ref{fig:lirhist}b is likely due
to the sparse data used to establish the two power-law relation for
galaxies with $1\times 10^{11}\la L_{\rm IR}\approx \la 5\times
10^{11}$~L$_{\odot}$ (see Figure~5d of \citealt{elbaz02}).  This range
of $L_{\rm IR}$ happens to encompass the typical IR luminosity of
$z\sim 2$ galaxies as inferred from X-ray and dust-corrected UV
estimates \citep{reddy04} and it is partly for this reason that we
favored our simple constant conversion relationship.

Regardless of the conversion used, we find that the bulk of the
galaxies in the $\ugr$ and $\bzk$/SF samples and detected at
$24$~$\mu$m have inferred infrared luminosities comparable to those of
local luminous infrared galaxies (LIRGs), with $10^{11}\la L_{\rm
IR}\la 10^{12}$~L$_{\odot}$.  Galaxies in the $\ugr$ sample with
$f_{\rm 24\mu m}\ga 8$~$\mu$Jy (corresponding to the $3$~$\sigma$
sensitivity limit) have $\langle L_{\rm IR}\rangle \sim 3.1\times
10^{11}$~L$_{\odot}$ for the constant conversion and $\langle L_{\rm
IR}\rangle \sim 2.1\times 10^{11}$~L$_{\odot}$ for the two power-law
conversion (the two power-law distribution is broader than that
obtained using the constant conversion).  Both the $\ugr$ and
$\bzk$/SF samples also host galaxies which, based on their inferred
$L_{\rm IR}$, would be considered ultra-luminous infrared galaxies
(ULIRGs) with $L_{\rm IR}\ga 10^{12}$~L$_{\odot}$.  Note that if we
excluded all direct X-ray detections, including the submillimeter
sources, the maximum inferred $L_{\rm IR}$ of $\ugr$ and $\bzk$/SF
galaxies is $\approx 2.5\times 10^{12}$~L$_{\odot}$, an infrared
luminosity which is similar to the detection limit of the {\it
Chandra} 2~Ms data for a galaxy at $z\sim 2$ assuming the
\citet{ranalli03} conversion between X-ray and FIR luminosity.

The $\bzk$/SF sample distribution shown in Figure~\ref{fig:lirhist}
includes galaxies that do not satisfy the $\ugr$ criteria (i.e.,
$\bzk$/SF--$\ugr$ galaxies).  These galaxies (to $\ks=21$) have a mean
IR luminosity that is identical to that of $\ugr$ galaxies to
$\ks=21$.  The average IR luminosity of $\ugr$ galaxies is $\sim 1.8$
times fainter than $\bzk$ galaxies since the $\ugr$ sample includes
galaxies which extend to fainter $\ks$ magnitudes than those in the
$\bzk$ sample.  Therefore, while the $\bzk$/SF--$\ugr$ galaxies have
redder $G-{\cal R}$ colors than required to satisfy the $\ugr$
criteria, it appears that their infrared luminosities are
still comparable to those of $\ugr$ galaxies (see also the discussion
in \S~\ref{sec:attenuation}), a result consistent with that obtained
in X-ray stacking analyses \citep{reddy05a}.  Figure~\ref{fig:lirhist}
indicates the DRG galaxies with photometric redshifts between
$1.5<z<2.6$ also span a large range in $L_{\rm IR}$, from luminosities
characteristic of LIRGs to ULIRGs.  We find a luminosity distribution
of DRGs to $\ks=21$ that is in good agreement with the $L_{\rm IR}$
distribution found by \citet{papovich05} for DRGs (to approximately
the same depth) in the GOODS-South field\footnote{The DRG sample of
  \citet{papovich05} extends to $\ks=23.2$ in AB magnitudes, or
  $\ks\sim 21.4$ in Vega magnitudes, over an area twice as large as
  studied here.}.  We note that $\ugr$ galaxies and DRGs to $\ks=20$
have the same inferred $L_{\rm IR}$ as $\ks<20$ galaxies selected in
other ways (e.g., using the $\bzk$/SF criteria).

The inferred $L_{\rm IR}$ for the one $\bzk$/PE selected galaxy with
$z\sim 1.5$ is $\sim 1.2\times 10^{12}$~L$_{\odot}$.  The mean $f_{\rm
  24\mu m}$ flux of MIPS-detected (and non X-ray detected) $\bzk$/PE
galaxies without redshifts is $\langle f_{\rm 24\mu m}\rangle \approx
81.4$~$\mu$Jy which, at the mean redshift of $\bzk$/PE galaxies (e.g.,
\citealt{reddy05a}; \citealt{daddi04b}) of $z\sim 1.7$, corresponds to
$L_{\rm IR} \sim 6\times 10^{11}$~L$_{\odot}$.  The $24$~$\mu$m
detection rate ($\sim 50\%$; Table~\ref{tab:stats}) of non X-ray
detected $\bzk$/PE galaxies implies some contamination by star-forming
galaxies; this is not unexpected given that photometric scattering can
have a significant effect on samples constructed by color selection
techniques \citep{reddy05a}.

\begin{figure}[hbt]
\plotone{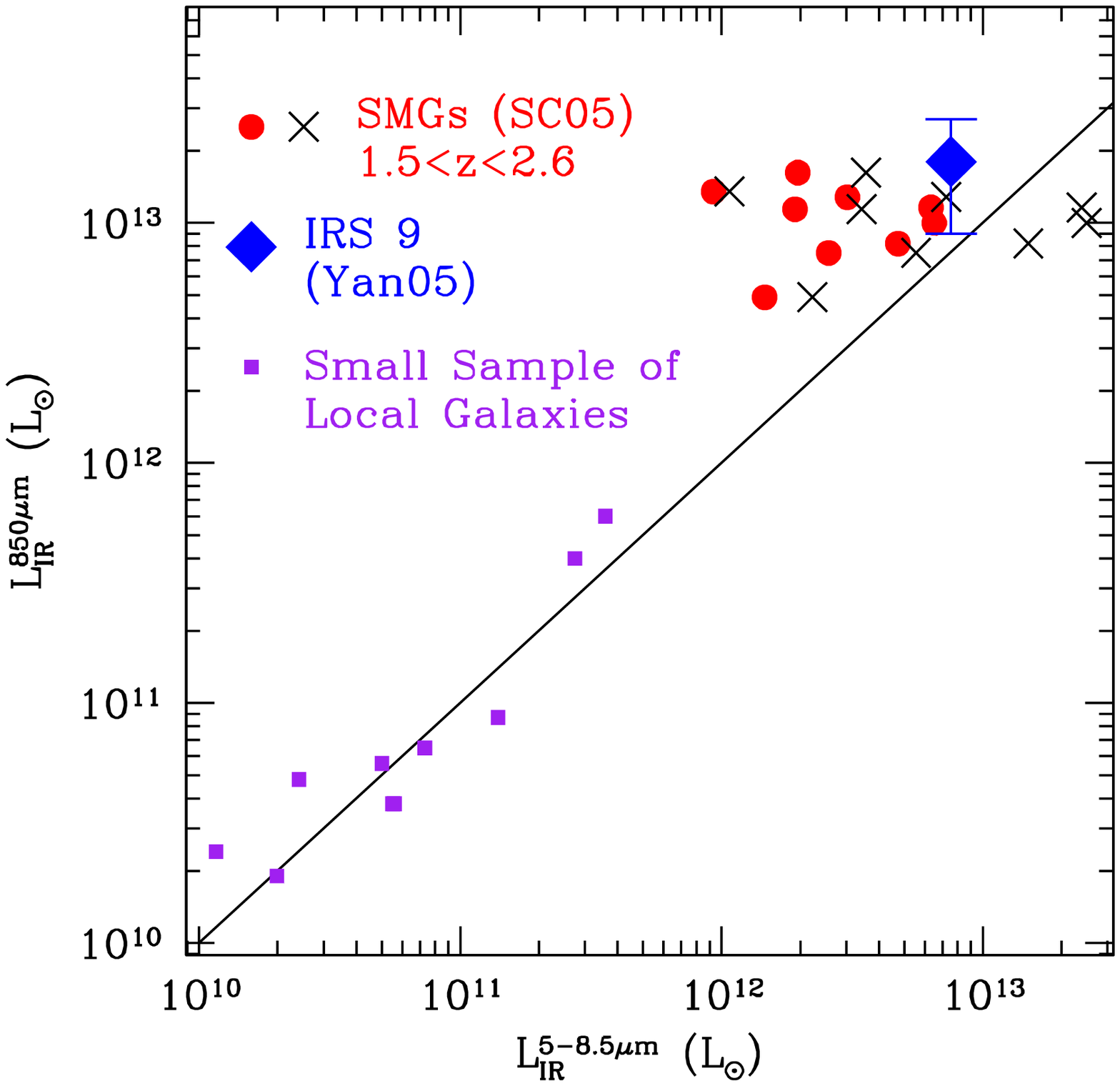}
\caption{$L_{\rm IR}^{850\mu m}$, inferred from the submillimeter
fluxes of radio-detected SMGs \citep{chapman05}, plotted against
$L_{\rm IR}^{5-8.5\mu m}$ inferred from $L_{5-8.5\mu m}$ assuming a
IR/MIR flux conversion of $17.2$ (solid circles) and the two power-law
conversion of \citet{elbaz02} (crosses).  The infrared luminosity of
IRS9 is inferred from MIPS $24$, $70$, and $160\mu m$ data and MAMBO
millimeter measurements \citep{yan05}.  Also shown are the local
star-forming templates listed in Table~\ref{tab:templates}.\\
\label{fig:lirsmg}}
\end{figure}

The radio-detected submillimeter galaxies to $S_{\rm 850\mu m}\sim
4.2$~mJy analyzed here have inferred $L_{\rm IR}$ of $10^{12}\la
L_{\rm IR}\la 10^{13}$~L$_{\odot}$, which can be directly compared
with their bolometric luminosities calculated based on the
submillimeter data presented by \citet{chapman05}.  The
$850$~$\mu$m-inferred bolometric luminosities ($L^{850\mu m}_{\rm
  IR}$) are sensitive to the assumed characteristic dust temperature
associated with a greybody fit to the dust SED and the emissivity.
For example, a change in the assumed dust temperature from $T_{\rm
  d}=36$~K (the median value for the sample of radio-detected SMGs in
\citealt{chapman05}) to a cooler temperature of $T_{\rm d}=31$~K
results in a factor of $\sim 5$ decrease in the inferred FIR
luminosities.  Figure~\ref{fig:lirsmg} shows the comparison between
$850$~$\mu$m and $24$~$\mu$m inferred bolometric luminosities of
radio-detected SMGs.  We also show the point corresponding to IRS9
from the \citet{yan05} sample of hyperluminous $z=2$ sources with IRS
spectroscopy --- this source has independent constraints on its FIR
luminosity based on MIPS $70$ and $160$~$\mu$m and MAMBO millimeter
measurements.  The infrared luminosity of IRS9 is $L_{\rm IR}\sim
1.8\times 10^{13}$~L$_{\odot}$ (constrained to within a factor of
$2-3$) based on these multi-wavelength measurements \citep{yan05}.

Figure~\ref{fig:lirsmg} shows that the IR/MIR flux ratio for IRS9 is
comparable (to within $\sim 1$~$\sigma$) to those of the local
star-forming galaxies listed in Table~\ref{tab:templates}--- these
local galaxies are $1-3$ orders of magnitude fainter than IRS9.
Judging the validity of our conversion for the hyper-luminous galaxies
at $z\sim 2$ is difficult given that very few of these galaxies have
independent constraints on their bolometric luminosities.  On the
other hand, the submillimeter ($850$~$\mu$m) inferred infrared
luminosities of bright SMGs are systematically a factor of $2-10$
higher than predicted using our conversion between $L_{\rm 5-8.5\mu
m}$ and $L_{\rm IR}$.  The crosses in Figure~\ref{fig:lirsmg}
demonstrate that the systematic offset cannot be completely accounted
for if we assume a two power-law conversion between $L_{\rm 5-8.5\mu
m}$ and $L_{\rm IR}$ --- there are still $4$ of $9$ SMGs with
$L^{850\mu m}_{\rm IR}$ that are factor of $2-10$ larger than
predicted from their $24$~$\mu$m fluxes, and the distribution of SMG
points when considering the two-power law conversion is not symmetric
about the line of equality (solid line in Figure~\ref{fig:lirsmg}).
One possibility is that the submillimeter estimates are in fact
correct and that our assumed conversion between mid-IR and IR
luminosities does not apply to SMGs.  The IRS sample considered here
consists of just one hyper-luminous galaxy at $z=2$, and if we ignore
this galaxy then the systematic offset of SMGs may indicate a
breakdown of our assumed conversion for the most luminous sources at
redshifts $z\sim 2$ with $L_{\rm IR}\ga 10^{13}$~L$_{\odot}$.  The
second possibility is that the submillimeter estimates overpredict the
infrared luminosities of SMGs and that our MIR-to-IR conversion is
correct.  This may not be surprising since the conversion between
submillimeter flux and bolometric luminosity is very sensitive to the
assumed dust temperature, and a decrease in the assumed temperature of
just a few degrees can reduce the inferred bolometric luminosity by a
factor of $\sim 5-10$ (see example above).  Finally, it is possible
that neither the submillimeter or mid-IR inferred infrared
luminosities of bright SMGs is correct.  We note that it is common for
these luminous galaxies to host AGN, and this can alter the observed
mid-IR and IR fluxes beyond what would be expected given pure star
formation (e.g., \citealt{armus04}; \citealt{fadda02};
\citealt{almaini99}; \citealt{fabian99}).  As another example, Arp 220
has anomalously low PAH emission for its bolometric luminosity (when
compared with other ULIRGs), suggesting that the galaxy contains a
buried quasar and/or a heavily dust-enshrouded starburst such that the
extinction at rest-frame $7$~$\mu$m is no longer negligible (e.g.,
\citealt{haas01}; \citealt{charmandaris97}).  {\it Spitzer} IRS
observations of bright radio-detected SMGs will be useful in
quantifying the relationship between the $5-8.5$~$\mu$m and bolometric
luminosities of these ultraluminous sources.

\begin{figure*}[hbt]
\plottwo{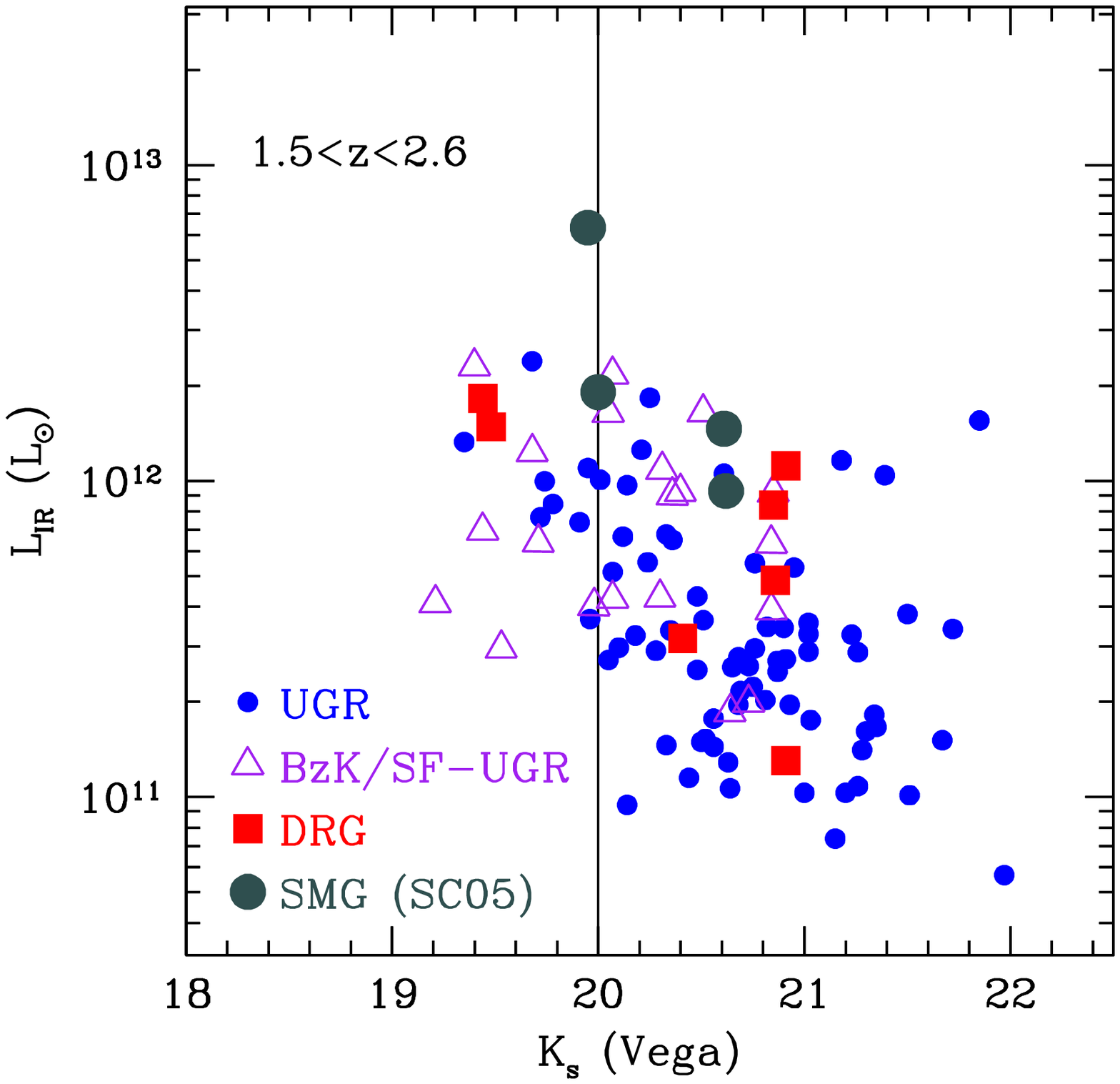}{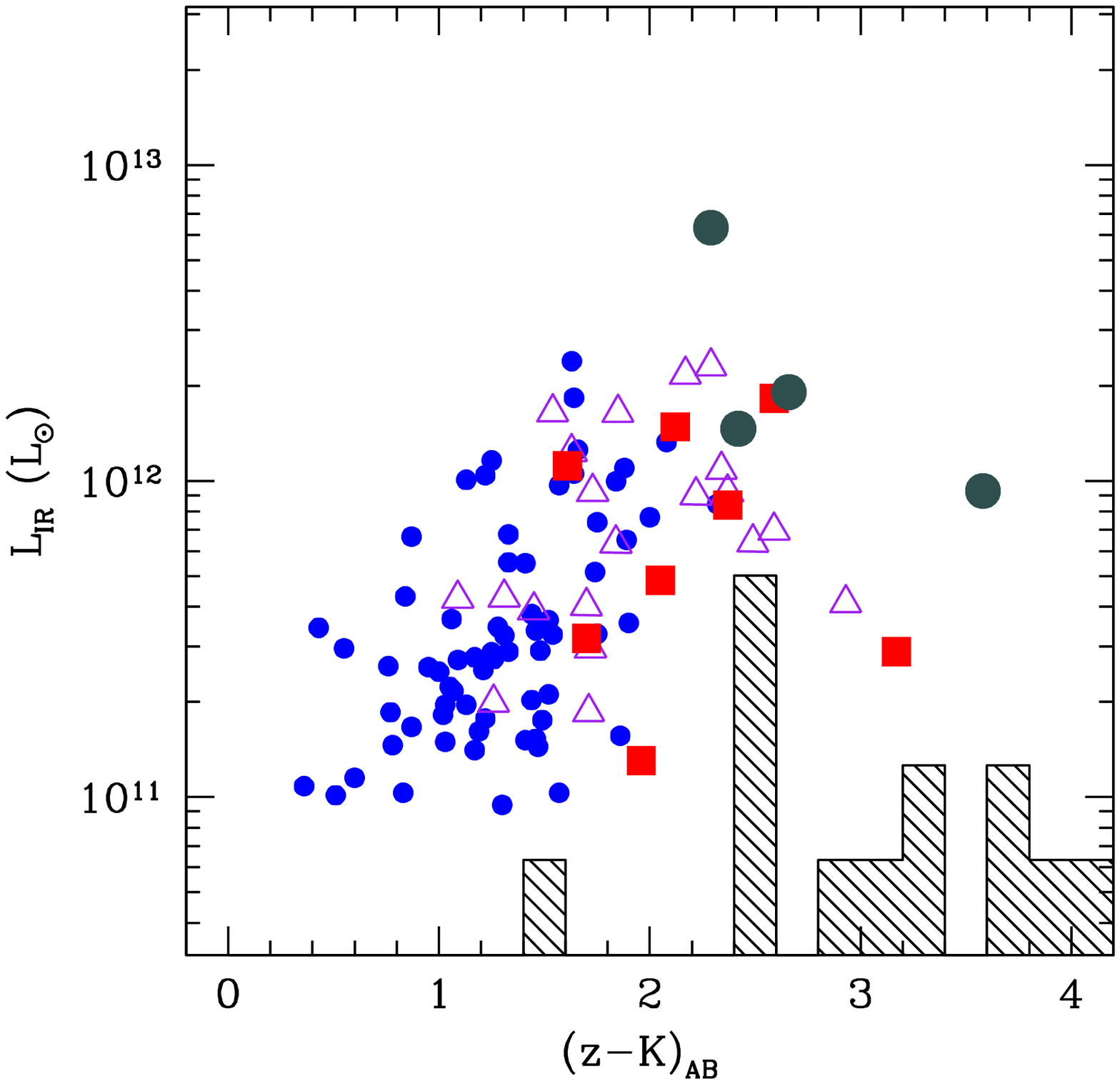}
\caption{{\it Left panel:} Distribution of $L_{\rm IR}$ as a function
of $\ks$ for galaxies in the various samples.  We have assumed a
IR/MIR flux ratio of $17.2$ for all galaxies including the
radio-detected SMGs.  {\it Right panel:} $L_{\rm IR}$ as a function of
$\zmk$ color.  Symbols are the same as in the left panel.  The
arbitrarily normalized histogram indicates the distribution in $\zmk$
color of DRGs and $\bzk$/PE galaxies undetected at $24$~$\mu$m.\\
\label{fig:lircol}}
\end{figure*}

A relevant line of investigation is to determine what the various
optical and near-IR color and magnitude selections imply for the
infrared luminosity distributions of the galaxies they select.
Figure~\ref{fig:lircol}a shows the inferred $L_{\rm IR}$ of galaxies
with redshifts $1.5<z<2.6$ as a function of $\ks$ magnitude.  We have
assumed the IR/MIR flux ratio of $17.2$ for the radio-detected SMGs
shown in the figure.  Galaxies with $\ks<20$ (e.g., K20 samples:
\citealt{cimatti02b}; \citealt{cimatti02a}) have $\langle L_{\rm
IR}\rangle\sim (1-2)\times 10^{12}$~L$_{\odot}$, similar to the value
found by \citet{daddi05b} for $\ks<20$ $\bzk$ selected galaxies in the
GOODS-N field.  Alternatively, we find $\langle L_{\rm IR}\rangle\sim
5\times 10^{11}$~L$_{\odot}$ for $\ugr$, $\bzk$, and DRG galaxies with
$20<\ks<21.0$.  As stated in \S~\ref{sec:f24toir}, the uncertainties
on any individual value of $\log L_{\rm IR}$ are $0.3$~dex for $\ugr$
galaxies with spectroscopic redshifts and $0.6$~dex for near-IR
selected ($\bzk$; DRG) galaxies with photometric redshifts.  At any
given $\ks$ magnitude, the range in $L_{\rm IR}$ spans an order of
magnitude assuming our constant conversion and larger than an order of
magnitude assuming the two power-law conversion of \citet{elbaz02}.
Finally we note that galaxies with $\ks<20$ at $z\sim 2$ which show
some signature of star formation (i.e., those that are MIPS detected)
generally have infrared luminosities that are a factor of two larger
than those of galaxies with $20<\ks<21$.  As discussed elsewhere,
there is also a population of massive galaxies with little detectable
star formation (e.g., \citealt{vandokkum04}; \citealt{reddy05a}).

We investigate this quiescent population of massive galaxies in the
context of their star-forming counterparts by examining $L_{\rm IR}$
as a function of $\zmk$ color (Figure~\ref{fig:lircol}b).  The $\zmk$
color probes the Balmer and $4000$~\AA\, breaks for galaxies at the
redshifts of interest here, and is also sensitive to the current star
formation rate (e.g., \citealt{reddy05a}; \citealt{daddi04b}).
Figure~\ref{fig:lircol}b shows that galaxies with redder $\zmk$ color
(up to $\zmk\sim 3$) have higher inferred $L_{\rm IR}$ (and larger
SFRs if the bolometric luminosity is attributed to star formation) on
average than galaxies with bluer $\zmk$ colors, a trend similar to
that found when stacking X-ray data \citep{reddy05a}.  A more
interesting result is indicated by the histogram which shows the
distribution in $\zmk$ color of DRGs and $\bzk$/PE galaxies that are
undetected at $24$~$\mu$m.  Of the 13 MIPS-undetected DRGs and
$\bzk$/PE galaxies, 7 have $\zmk>3$.  \citet{reddy05a} found DRGs with
$\zmk\ga 3$ to have little X-ray emission and had colors similar to
those of IRAC Extremely Red Objects (IEROs; \citealt{yan05}).  The
lack of $24$~$\mu$m detections for these red $\zmk$ sources further
supports the notion that they have little current star formation.  It
also rules out the possibility that they harbor Compton-thick obscured
AGN as an explanation for their lack of X-ray emission, since we would
then expect them to be bright at $24$~$\mu$m.

\subsection{Stacked $24$~$\mu$m Flux}
\label{sec:f24stack}

\begin{figure}[hbt]
\plotone{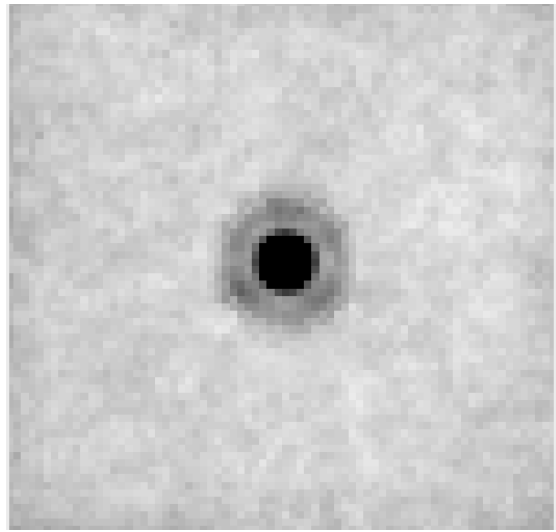}
\caption{Stacked $24$~$\mu$m flux of 48 isolated MIPS-undetected
$\ugr$ galaxies ($f_{\rm 24\mu m}\la 8$~$\mu$Jy) with spectroscopic
redshifts $1.5<z<2.6$, indicating a mean flux per
undetected galaxy of $\langle f_{24\mu m}\rangle \approx
3.30\pm0.48$~$\mu$Jy.\\
\label{fig:ugrf24stack}}
\end{figure}

The high quality and depth of the MIPS data enable us to study the
average properties of galaxies that are (individually) undetected at
$24$~$\mu$m by stacking their emission.  The data were stacked by
considering all galaxies undetected at $24$~$\mu$m and X-ray
wavelengths.  To avoid contaminating the stacked signal, we only added
galaxies to the stack if there were no bright MIPS sources within
$\approx 5\arcsec$ of those galaxies.  To ensure a clean signal, we
extracted sub-images around each undetected galaxy, subtracted all
{\it detected} point sources within those sub-images using the
empirical PSF, and added the sub-images together.  The stacked flux
was measured in a manner similar to the measurement of detected MIPS
sources.  Figure~\ref{fig:ugrf24stack} shows the stacked image of $48$
MIPS-undetected $\ugr$ galaxies with spectroscopic redshifts
$1.5<z<2.6$.  The mean flux per object is $\langle f_{\rm 24\mu
m}\rangle\approx 3.30 \pm 0.48$~$\mu$Jy, where the error is the
dispersion of the background in the {\it stacked} image.  At a mean
redshift of $\langle z\rangle =2.05$, this flux corresponds to $L_{\rm
IR} \approx 2\times 10^{10}$~L$_{\odot}$.  Combining this result with
the mean $L_{\rm IR}$ of MIPS-detected $\ugr$ galaxies implies a mean
across the entire sample, neglecting ``confused'' sources, of $\approx
2.3\times 10^{11}$~L$_{\odot}$.  This mean value does not change
significantly if we add in directly-detected X-ray sources (including
radio-detected SMGs to $S_{\rm 850\mu m}\sim 5$~mJy) because of their
small number compared to the typical (less luminous) $\ugr$ galaxy.
The mean value of $L_{\rm IR}\approx 2.3\times 10^{11}$~L$_{\odot}$ is
in excellent agreement with the average of $L_{\rm IR}\approx 3\times
10^{11}$~L$_{\odot}$ found from stacked X-ray/radio and dust corrected
UV estimates \citep{reddy04}.  This suggests that the non-detection of
galaxies at $24$~$\mu$m is due to them having lower SFRs and not because
they are deficient in PAH emission for a given $L_{\rm IR}$.  The
advantage of the $24$~$\mu$m data over X-ray/radio data is that we can
estimate bolometric luminosities for individual $L^*$ (LIRG) galaxies
at $z\sim 2$ rather than ensembles of galaxies.

Combining our estimate of the MIPS-inferred average IR luminosity of
$\ugr$ galaxies with the stacked radio results of \citet{reddy04}, we
find that the radio-IR relation appears valid on average for the
sample.  To quantify the radio-IR ratio for the sample, we compute the
``q'' parameter \citep{condon91}:
\begin{eqnarray}
q\equiv\log \bigl({{\rm FIR} \over {\rm 3.75\times10^{12}~W~m^{-2}}}\bigr) -
\log \bigl({S_{\rm 1.4~GHz} \over {\rm W~m^{-2}~Hz^{-1}}}\bigr),
\end{eqnarray}
where $S_{\rm 1.4~GHz}$ is the rest-frame $1.4$~GHz flux density in units
of W~m$^{-2}$~Hz$^{-1}$ and 
\begin{eqnarray}
{\rm FIR} \equiv 1.26\times10^{-14} (2.58S_{\rm 60~\mu m}+S_{\rm
100~\mu m})~{\rm W~m^{-2}},
\end{eqnarray}
where $S_{\rm 60~\mu m}$ and $S_{\rm 100~\mu m}$ are the IRAS $60$ and
$100$~$\mu$m flux densities in Jy \citep{helou88}.  The implied ``q''
value for the $\ugr$ sample is $\langle q\rangle \sim 2.5$ if we
assume $\log [S_{\rm 60~\mu m}/S_{\rm 100~\mu m}]\sim 0.2$.  This
value of $q$ is in excellent agreement with the value of $q\sim 2.4$
found for $\ga 10^{11}$~L$_{\odot}$ galaxies in the IRAS 2~Jy sample
\citep{yun01}.  We also stacked the $24$~$\mu$m data for undetected
$\bzk$/PE and DRG galaxies in the same manner as described above,
which yielded a mean flux per object of $\langle f_{\rm 24\mu
  m}\rangle \sim 2.72\pm 1.65$~$\mu$Jy.  As noted in
\S~\ref{sec:lirdist}, most of these sources have very red $\zmk$
colors, and their low-level mid-IR and X-ray emission indicate they
have low SFRs.  Galaxies with $f_{\rm 24\mu m}\la 8$~$\mu$Jy are
discussed further below.

\section{Dust Attenuation in Optical and Near-IR Selected Galaxies}
\label{sec:attenuation}

Aside from inferring the infrared luminosity distributions, we can use
the MIPS data to examine the relationship between dust extinction and
rest-frame UV spectral slope.  \citet{meurer99} found a relation
between the rest-frame UV spectral slope, $\beta$, and the attenuation
of local UV-selected starburst galaxies, parameterized by the ratio
$L_{\rm FIR}/L_{\rm 1600}$, where $L_{\rm 1600}$ is the rest-frame
$1600$~\AA\, luminosity uncorrected for extinction.  This relation
appears to fail, however, for the most luminous starbursts such as
ULIRGs \citep{goldader02} and radio-detected SMGs \citep{chapman05}.
A greater proportion of the star formation in galaxies with $L_{\rm
IR}\ga 10^{12}$~L$_{\odot}$ will be obscured by dust as compared with
LIRG-type starbursts (e.g., see \S~\ref{sec:dustobs}).  Therefore,
whatever UV emission is able to escape from the optically-thin regions
of ULIRGs will constitute a lower fraction of the total bolometric
luminosity.  As a result, the rest-frame UV light can substantially
underpredict (by a factor of $10-100$) the bolometric luminosities of
the most luminous starbursts, such as radio-detected SMGs
\citep{chapman05}.  Normal (``quiescent'') star forming galaxies also
appear to deviate from the \citet{meurer99} relation, such that
$L_{\rm FIR}/L_{\rm 1600}$ is lower for a given amount of UV reddening
than in starburst galaxies (e.g, \citealt{laird05}; \citealt{bell02})
a result that may be tied to the varying ratio of current to
past-average star formation rate of normal star forming galaxies
\citep{kong04}.  Alternatively, the star formation in local quiescent
galaxies (i.e., those with low SFRs) is more distributed than in local
starbursts so that a failure of the starburst reddening law may
reflect a different distribution of dust with respect to the star
forming regions in quiescent galaxies.  Observations of radio-detected
SMGs and quiescently star-forming galaxies suggests that the
\citet{meurer99} and \citet{calzetti00} laws do not apply to these
sources.

The rest-frame UV spectral and mid-IR data of $z\sim 2$ galaxies allow
us to investigate how well the high redshift galaxies follow the local
dust attenuation relation.  The full SED modeling of $\ugr$ galaxies
in the GOODS-N field yields estimates of the best-fit star formation
history ($\tau$), age, mass, SFR, and $\ebmv$ color excess for each
galaxy (\citealt{erb06b}; \citealt{shapley05}).  The mean fractional
uncertainty in $\ebmv$ is $\langle \sigma_{\rm \ebmv}/\ebmv\rangle =
0.7$ as determined from Monte Carlo simulations.  To convert $\ebmv$
to $\beta$ we assumed that $1$~mag of extinction at $1600$~\AA\,
($A_{\rm 1600}=1$) corresponds to $\ebmv\approx 0.092$ (e.g.,
\citealt{calzetti00}).  For most galaxies, the best-fit star formation
history is close to that of a constant star formation history (with
decay time-scale $\tau=\infty$).  The most massive galaxies at $z\sim
2$ (with stellar masses $M^{\ast} \ga 10^{11}$~M$_{\odot}$) are generally
better fit with declining star formation histories.  We have assumed a
CSF model for galaxies unless such a model provides a much poorer fit
to the observed data than a declining star formation history.

\subsection{Results for Optically Selected Galaxies}
\label{sec:optical}

\begin{figure*}[hbt]
\plottwo{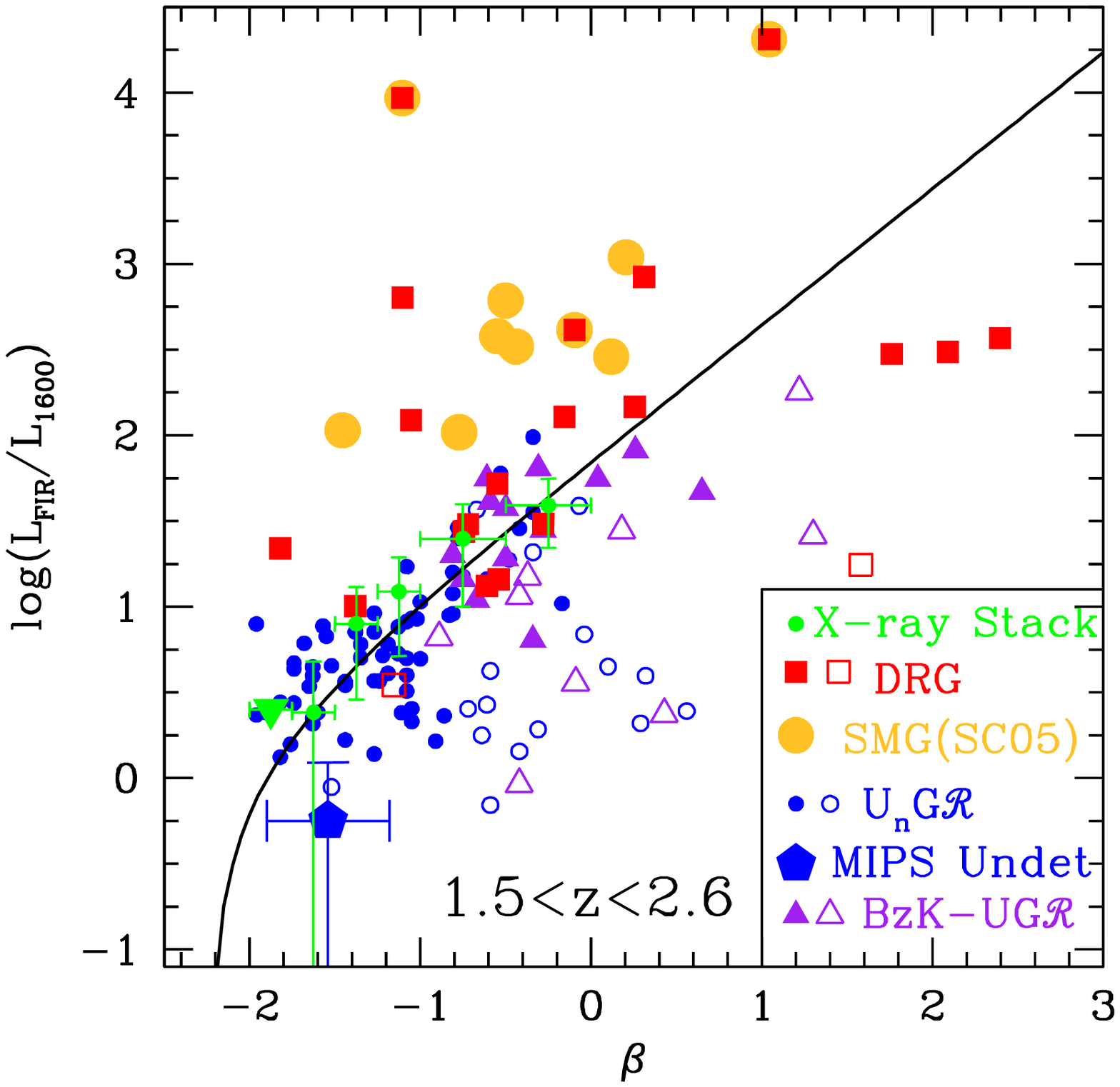}{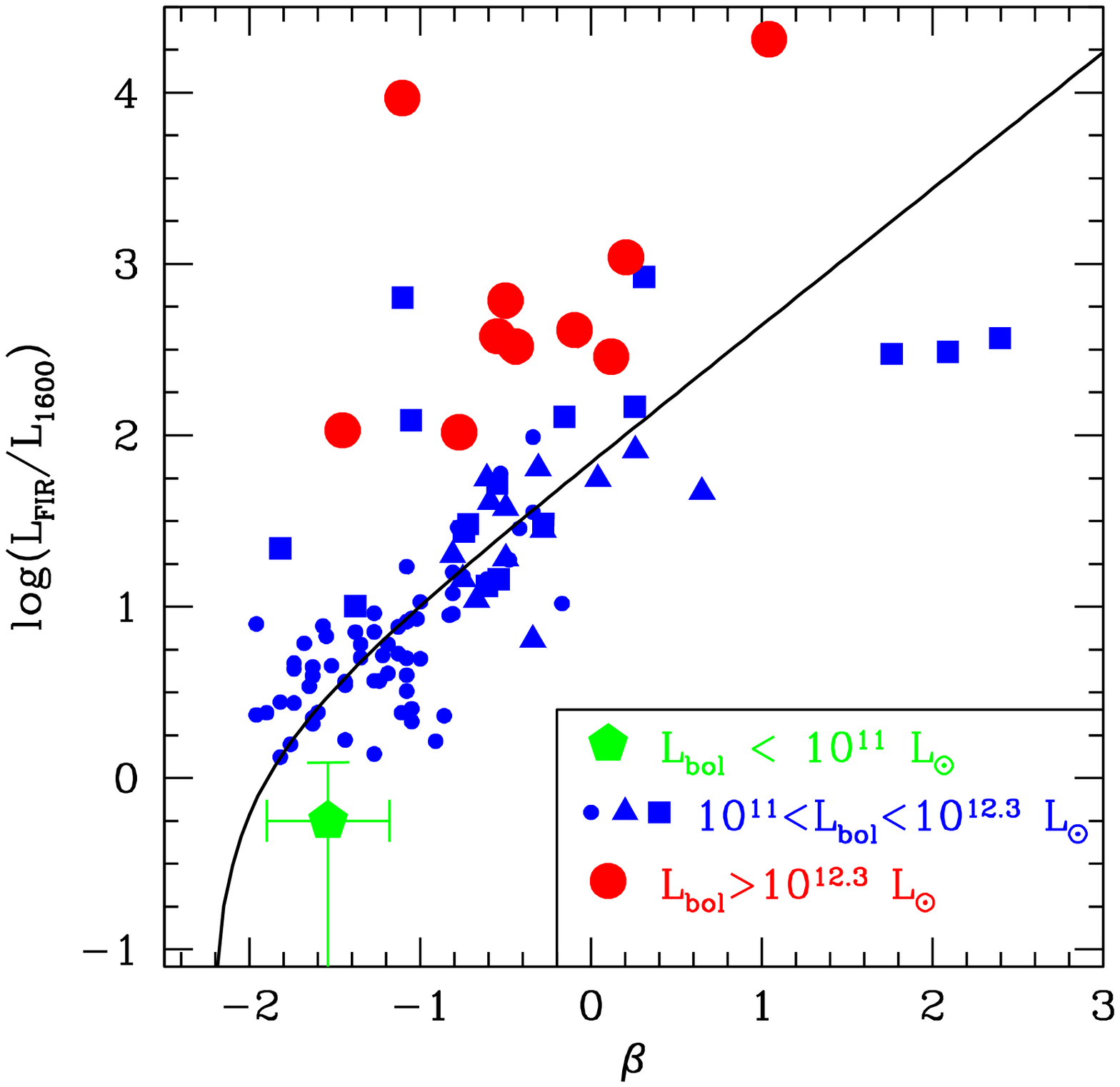}
\caption{({\it Left}) Dust absorption, parameterized by $F_{\rm
FIR}/F_{1600}$, versus rest-frame UV spectral slope, $\beta$, for
galaxies with redshifts $1.5<z<2.6$.  Filled and open symbols
respectively denote galaxies with inferred ages of $>100$~Myr and
$<100$~Myr.  The large pentagon shows the results for $\ugr$ galaxies
undetected at $24$~$\mu$m, using the stacked results of
\S~\ref{sec:f24stack}.  The horizontal and vertical error bars on this
point reflect the dispersion about the mean $\beta$ and mean $L_{\rm
FIR}/L_{\rm 1600}$ of $24$~$\mu$m undetected sources.  The green
points represent the results from an X-ray stacking analysis as
described in the text.  The solid line indicates the \citet{meurer99}
relation found for local UV-selected starburst galaxies. ({\it Right})
Same as {\it left} panel, but excluding galaxies younger than
$100$~Myr and color-coding objects by $L_{\rm bol}$.\\
\label{fig:pahuvebmv}}
\end{figure*}

Figure~\ref{fig:pahuvebmv} shows $L_{\rm FIR}/L_{\rm 1600}$ versus
$\beta$ for spectroscopically-confirmed $\ugr$ galaxies with redshifts
$1.5<z<2.6$.  The FIR luminosity is computed from $L_{\rm 5-8.5\mu m}$
using the procedure described in \S~\ref{sec:f24toir}.  We estimated
the rest-frame $1600$~\AA\, luminosity from either the $U_{\rm n}$,
$G$, or ${\cal R}$ magnitude depending on the redshift of the galaxy.
The majority of $\ugr$ galaxies with inferred ages $\ga 100$~Myr
(solid circles) appear to agree well with \citet{meurer99} relation,
shown by the solid curve\footnote{Assuming the two power-law
conversion to compute $L_{\rm FIR}$ results in a similar distribution
of points around the \citet{meurer99} relation but with larger
scatter.}.  

$\ugr$ galaxies with the youngest inferred ages ($\la 100$~Myr; open
circles) exhibit a large offset from the \citet{meurer99} relation in
the sense that they exhibit redder UV colors for a given dust
obscuration than older galaxies which do follow the relation.  Note
that we have assumed a CSF model for the young galaxies shown in
Figure~\ref{fig:pahuvebmv}.  The inferred ages of these galaxies are
typically smaller than $50$~Myr, which is approximately the dynamical
time across the galaxy.  Assuming such small (and unrealistic ages)
will cause us to overestimate $\ebmv$ for these sources.  The change
in $\ebmv$ that results from fixing the age of the young galaxies to
$100$~Myr ($\Delta\ebmv = 0.09$) is not enough to completely account
for the offset of the young galaxies from the \citet{meurer99}
relation.  This suggests that \citet{calzetti00} law may not be
applicable to these young galaxies because of a different
configuration of dust with respect to the star-forming regions, in
which case a grayer extinction law may be appropriate.  As one
example, the well-studied lensed Lyman Break Galaxy MS1512-cB58 has an
inferred age of $\sim 70-100$~Myr and millimeter continuum
observations suggest that its infrared luminosity is smaller than one
would predict from its UV reddening \citep{baker01}.  Regardless of
the assumed extinction law, these young galaxies in the samples
examined here have similar bolometric (sum of observed IR and UV)
luminosities as older galaxies (see \S~\ref{sec:dustobs}).

The deep X-ray data in the GOODS-N field allow us to estimate (X-ray
inferred) average infrared luminosities for well-defined samples of
galaxies (e.g., \citealt{reddy05a}; \citealt{laird05};
\citealt{nandra02}; \citealt{brandt01}).  The green points in
Figure~\ref{fig:pahuvebmv} show the expected dust absorption inferred
from the X-ray data as a function of $\beta$.  These points were
determined by stacking the X-ray data for non X-ray detected $\ugr$
galaxies (with ages $>100$~Myr) in bins of $\beta$.  We only
considered stacking galaxies with ages $>100$~Myr since the X-ray
emission is sensitive to the star formation rate once O and B stars
evolve to produce high mass X-ray binaries, which is roughly
$10^{8}$~years after the onset of star formation.  The average X-ray
flux per bin was converted to a FIR flux using the \citet{ranalli03}
relation.  Dividing the average FIR flux per bin by the average
$1600$~\AA\, luminosity of objects in each bin yields the green
points.  The horizontal error bars indicate the bin width and the
vertical error bars indicate the dispersion in stacked X-ray flux
(computed using the procedure described in \citet{reddy05a} and
\citet{reddy04}) added in quadrature with the dispersion in the
FIR/X-ray luminosity relation found by \citet{ranalli03}.  The X-ray
results reproduce the \citet{meurer99} relation very well, providing
an independent confirmation that typical $\ugr$ galaxies abide by the
local dust obscuration relation for starburst galaxies.  The X-ray and
mid-IR data indicate that the UV slope can be used to deduce the
extinction corrections for these typical galaxies and that such
galaxies have UV light that is only moderately extinguished in most
cases.  The agreement between the MIPS inferred FIR luminosities and
X-ray inferred FIR luminosities (obtained with the
empirically-determined \citet{ranalli03} relation) suggests that the
\citet{ranalli03} relation provides a reasonable description for most
of the galaxies considered here.  For comparison, adopting the
\citet{persic04} relation would yield FIR luminosities five times in
excess of those predicted from the $24$~$\mu$m fluxes for the vast
majority of $z\sim 2$ galaxies.

We compute the average dust obscuration of $\ugr$ galaxies undetected
at $24$~$\mu$m using the stacking results of \S~\ref{sec:f24stack},
and the result is denoted by the large pentagon in
Figure~\ref{fig:pahuvebmv}.  The stacked $24$~$\mu$m FIR luminosity of
MIPS undetected galaxies is divided by their average rest-frame
$1600$~\AA\, luminosity.  For these galaxies, $L_{\rm 1600}$ is on
average $1.6$ times larger than their inferred $L_{\rm FIR}$
indicating they are significantly less obscured than galaxies detected
at $24$~$\mu$m.  These undetected galaxies also have relatively blue
rest-frame UV SEDs (as indicated by their average $\beta$) compared to
$24$~$\mu$m detected galaxies.  Furthermore, the results of
Figure~\ref{fig:pahvxray} indicate these faint sources have lower
X-ray emission than $24$~$\mu$m detected galaxies.  All of these
observations combined suggest that galaxies are undetected at
$24$~$\mu$m because they have lower SFRs and are less obscured than
galaxies with brighter $24$~$\mu$m fluxes.  We further explore the
nature of these MIPS undetected sources in \S~\ref{sec:undet}.

\subsection{Results for Near-IR and Submillimeter Selected Galaxies}
\label{sec:near-ir}

Also shown in Figure~\ref{fig:pahuvebmv} are $\bzk$/SF galaxies to
$\ks=21$ not satisfying the $\ugr$ criteria.  As pointed out in
\S~\ref{sec:lirdist}, these $\bzk$/SF selected galaxies have inferred
$L_{\rm IR}$ that are comparable to those of $\ugr$ galaxies to the
same $\ks$ limit, but of course with redder $G-{\cal R}$ colors and a
${\cal R}\sim 0.5$~mag fainter on average than $\ugr$ galaxies to the
same $\ks$ limit.  The results of Figure~\ref{fig:pahuvebmv} suggest
that $\bzk$/SF galaxies lying outside the color range selected by the
$\ugr$ criteria also follow the \citet{meurer99} relation.  Similar to
the results found in \S~\ref{sec:optical} for most $\ugr$ galaxies,
the mid-IR data indicate that the UV light from most $\bzk$/SF
galaxies appears to be moderately extinguished and that the UV slope
can be used to estimate their attenuation.

Almost all of the radio-detected SMGs considered here have inferred
dust absorption factors (when we assume the $850$~$\mu$m-inferred
bolometric luminosities) that are at least a magnitude larger than
predicted by the \citet{meurer99} relation for a given rest-frame UV
slope.  The discrepancy is not as substantial (i.e., it is reduced by
a factor of $2-10$) if we inferred $L_{\rm IR}$ of the SMGs from their
$24$~$\mu$m fluxes assuming our conversion between MIR and IR
luminosity.  The $\ugr$ criteria are designed to select objects where
followup spectroscopy is feasible, and this usually implies setting a
limit to the allowed $\ebmv$ (or $\beta$) of objects in the sample.
However, given that at least half the galaxies with $L_{\rm bol}\ga
10^{12}$~L$\odot$ have UV slopes comparable to that of the {\it
typical} $\ugr$ galaxy, it is not uncommon to find such dust-obscured
galaxies in optical surveys.

Of the limited sample of DRGs with photometric redshifts $1.5<z<2.6$,
at least half lie above the local starburst attenuation law.  We are
able to find DRGs that agree with the \citet{meurer99} relation since
the MIPS data studied here are significantly deeper (by a factor of
$\sim 5$) than the data considered in \citet{papovich05}.  In
particular, we find the surface density of DRGs between $1.5<z<2.6$
with $1\la \log(F_{\rm FIR}/F_{\rm 1600})\la 2$ of $\ga
0.14$~arcmin$^{-2}$ (this is a lower limit since there are number of
DRGs without photometric redshifts, some of which may truly lie at
redshifts $1.5<z<2.6$), which is at least a factor of $20$ higher than
in \citet{papovich05}.  Our results suggest that the DRG population
consists of galaxies with a very wide range in star formation rate,
from galaxies with little or no star formation (DRGs with very red
$\zmk$ colors; \S~\ref{sec:lirdist}) to those which are heavily
obscured and rapidly forming stars.

\subsection{Relationship between $\beta$ and Obscuration 
as a Function of Luminosity}

Figure~\ref{fig:pahuvebmv}b shows galaxies with ages $>100$~Myr within
the samples, color-coded by their $L_{\rm bol}$.  Virtually all
objects with $L_{\rm bol}$ in the range $10^{11}<L_{\rm
bol}<10^{12.3}$~L$_{\odot}$ have $\beta$ which appear to reproduce
their obscuration as inferred from the \citet{meurer99} and
\citet{calzetti00} laws.  There is some weaker evidence that the
galaxies with the lowest SFRs (undetected at $24$~$\mu$m) as indicated
by the green pentagon in Figure~\ref{fig:pahuvebmv}b follow a
different extinction law.  More pronounced, however, is the systematic
offset of the most luminous galaxies considered here with $L_{\rm
bol}>10^{12.3}$~L$_{\odot}$.  These ultraluminous galaxies have
rest-frame UV slopes that underpredict their obscuration by a factor
of $10-100$.  The main results of Figure~\ref{fig:pahuvebmv}b indicate
that the relationship between UV reddening and obscuration is strongly
dependent on the bolometric luminosity, but that most LIRG galaxies at
$z\sim 2$ follow the local relation.

\section{Relationship Between Dust Obscuration and Bolometric Luminosity}
\label{sec:dustobs}

\begin{figure*}[hbt]
\plotone{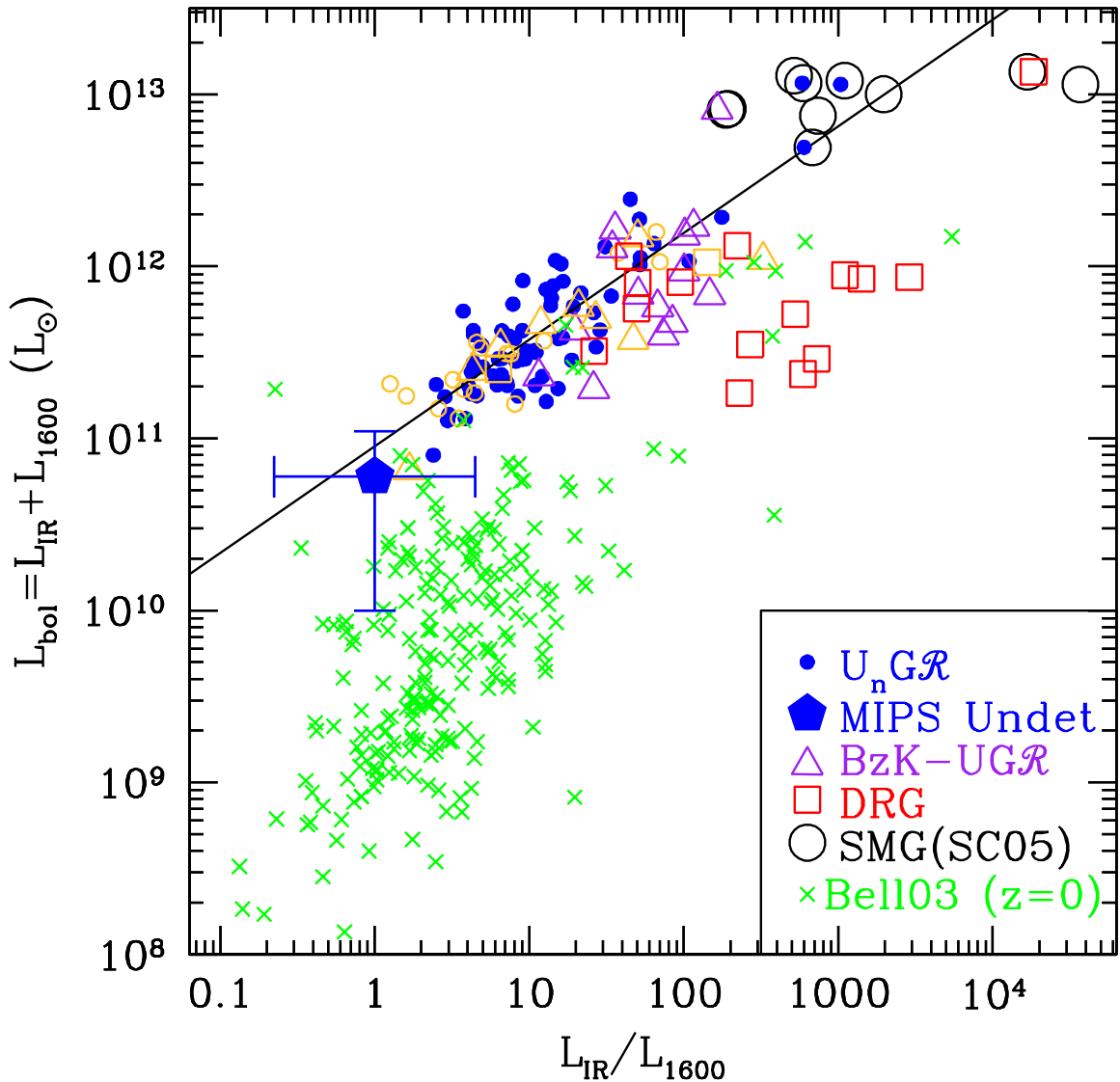}
\caption{Bolometric luminosity, approximated as the sum of the IR and
UV luminosities, versus dust obscuration.  Shown are the distributions
for $z\sim 2$ galaxies, and the solid line indicates the best-fit
linear relation for spectroscopically confirmed $\ugr$ galaxies
detected at $24$~$\mu$m.  For comparison, the pentagon denotes the
result for galaxies in the $\ugr$ sample undetected at $24$~$\mu$m,
and the yellow symbols denote objects with inferred ages $<100$~Myr.
We also show local galaxies from the compilation of \citet{bell03}.
Galaxies of a {\it given} bolometric luminosity are typically $8-10$
times less dust obscured at high redshift than at the present epoch.\\
\label{fig:bolo}}
\end{figure*}

The bolometric luminosity of star-forming galaxies can be
well-approximated by the sum of the IR and UV luminosities as
indicated in Equation~\ref{eq:bolo}.  Figure~\ref{fig:bolo} shows
$L_{\rm bol}$ as a function of dust obscuration for objects in the
various samples assuming a constant conversion relation between mid-IR
and total IR luminosity.  Typical (LIRG) galaxies at $z\sim 2$ will
have $L_{\rm bol} \approx L_{\rm IR}$ where $\sim 90\%$ of the
bolometric luminosity is emitted in the infrared.  The bolometric
luminosity is strongly correlated with dust obscuration: galaxies with
larger bolometric luminosities are more dust obscured than less
luminous galaxies.  The best-fit linear trend for
spectroscopically-confirmed $\ugr$ galaxies detected at $24$~$\mu$m is
\begin{eqnarray}
\log L_{\rm bol} = (0.62\pm0.06)\log {L_{\rm IR}\over 
L_{\rm 1600}} + (10.95\pm0.07)
\label{eq:bolo}
\end{eqnarray}
(solid line in Figure~\ref{fig:bolo}; we note that the two axes are
not independent of each and may partly account for the tight scatter
in the correlation).  $\ugr$ galaxies undetected at $24$~$\mu$m are
indicated by the pentagon.  These undetected galaxies have an average
bolometric luminosity of $\langle L_{\rm bol}\rangle\sim 6\times
10^{10}$~L$_{\odot}$ and have UV luminosities that are a factor of
$\sim 10$ less dust obscured than the typical $24$~$\mu$m detected
$\ugr$ galaxy.  Approximately half of the bolometric luminosity of
these $24$~$\mu$m undetected galaxies is emitted in the UV.  Galaxies
with inferred ages $<100$~Myr (yellow symbols in
Figure~\ref{fig:bolo}) have $L_{\rm bol}$ comparable to those of older
galaxies at $z\sim 2$, suggesting that these young galaxies have
similar IR/MIR ratios as older galaxies.  Therefore, the deviation of
the young galaxies from the \citet{meurer99} law as noted in
\S~\ref{sec:attenuation} suggests that we have over-estimated $\ebmv$
for these young sources and/or they may follow a steeper (e.g.,
SMC-like) extinction law.

We note that $\ugr$ galaxies with $\ks<21$ have bolometric
luminosities and dust obscuration comparable to $\bzk$ galaxies to
$\ks=21$ that are not optically selected.  This implies that objects
missed by optical selection but which appear in the near-IR selected
$\bzk$ sample are not missed because they are more dust obscured, a
result corroborated by X-ray stacking analyses \citep{reddy05a}.  The
bright radio-detected SMGs have the highest inferred bolometric
luminosities among all galaxies considered here, of order $\sim
10^{13}$~L$_{\odot}$ with dust obscuration factors $\ga 100$.
Galaxies in the $\ugr$, $\bzk$, and radio-detected SMG samples
detected at $24$~$\mu$m mostly follow the linear relation denoted by
the solid line in Figure~\ref{fig:bolo}.  For SMGs, $L_{\rm
bol}\approx L_{\rm IR}$, so assuming the submillimeter estimates of
$L_{\rm IR}$ (rather than the mid-IR estimates shown in
Figure~\ref{fig:bolo}) will move the SMGs in a direction parallel to
the $z\sim 2$ trend.

DRGs detected at $24$~$\mu$m span a large range in $L_{\rm IR}/L_{\rm
1600}$.  About half the DRGs follow the linear trend established for
optically selected galaxies at $z\sim 2$.  The remaining half of DRGs
have similar bolometric luminosities to those which follow the $z\sim
2$ trend, but the UV luminosities are a magnitude more attenuated than
what we would have predicted from the \citet{meurer99} law.  The SED
analysis (\S~\ref{sec:photoz}) demonstrates that all of the DRGs which
follow the $z\sim 2$ trend are all relatively young galaxies (ages
$\la 2$~Gyr) and have lower stellar masses ($M^{\ast}\la
10^{11}$~M$_{\odot}$).  In contrast, DRGs which are offset from the
trend are all older (ages $\ga 2$~Gyr) and all have masses $>1.2\times
10^{11}$~M$_{\odot}$.  The offset could be explained naturally if the
dust masses of galaxies increase as they age, a natural consequence of
star formation.  Note that if the massive, metal-rich DRGs have
stronger PAH flux for a given IR luminosity than the younger galaxies,
then this would serve to only increase the offset between the massive
DRGs and the $z\sim 2$ trend.  In fact, stacking the X-ray data for
the younger and older DRGs indicates they have very similar bolometric
luminosities, confirming the results obtained by inferring $L_{\rm
IR}$ from the PAH flux.  The results of Figure~\ref{fig:bolo} suggest
that much of the dust in galaxies with the largest stellar masses was
produced by star formation prior to the episode currently heating the
dust.  Therefore, such galaxies will have larger dust obscuration for
a given bolometric luminosity.  Assuming the two power-law conversion
of \citet{elbaz02} would result in a $z\sim 2$ trend with a slope
$20\%$ larger than given in Equation~\ref{eq:bolo}, but with
approximately the same intercept within the uncertainties, so our
conclusions would be unchanged.

As galaxies are enriched with dust as they age, then we expect to see
an even greater difference in dust obscuration between $z\sim 2$
galaxies and those at the present epoch.  To investigate this, we
examined $L_{\rm bol}$ versus $L_{\rm IR}/L_{\rm 1600}$ for the sample
of local galaxies compiled by \citet{bell03}, shown by crosses in
Figure~\ref{fig:bolo}.  The local sample includes the ULIRGs studied
by \citet{goldader02}.  Unfortunately, the UV and IR data for local
LIRG and ULIRGs are relatively sparse.  However, of the small sample
of local galaxies with $L_{\rm bol}\ga 10^{11}$~L$_{\odot}$, almost
all ($10$ of $11$) lie to the right of the linear trend at $z\sim 2$
and at least half occupy the same region as the old, massive DRGs at
$z\sim 2$.  In fact, an interesting corollary to the above discussion
is that massive, star-forming DRGs at $z\sim 2$ are more analogous to
local ULIRGs than bright SMGs at $z\sim 2$, both in terms of
bolometric luminosity {\it and} dust-obscuration.  Local ULIRGs
undoubtedly carry a significant amount of dust into their current star
formation episodes (e.g., \citealt{goldader02}), so it not surprising
that they have similar dust-obscuration factors as massive,
star-forming DRGs at $z\sim 2$.  On the other hand, most galaxies
lying on the $z\sim 2$ trend, including many bright SMGs, are likely
undergoing their first major episode of star formation and have
relatively low dust-to-gas ratios, unlike the more massive (offset)
DRGs and local galaxies.

The offset between $z=0$ and $z\sim 2$ galaxies can be seen at fainter
bolometric luminosities where the local sample includes more galaxies
($10^{10}\la L_{\rm bol}\la 10^{11}$).  Restricting our analysis to
galaxies in the \citet{bell03} sample with $L_{\rm bol}$ comparable to
those of $24$~$\mu$m undetected $z\sim 2$ galaxies, we find that the
local sample is on average $\sim 10$ times more dust obscured than
$24$~$\mu$m undetected galaxies at $z\sim 2$.  Further, recent GALEX
results indicate that local near-UV selected galaxies with $L_{\rm
bol}=10^{11}$~L$_{\odot}$ have a mean dust obscuration factor of
$\approx 10$; this is $8$ times larger than the inferred dust
obscuration of a $L_{\rm bol}=10^{11}$~L$_{\odot}$ galaxy at $z=2$
\citep{burgarella05}.  To summarize, the important result from
Figure~\ref{fig:bolo} is that galaxies of a {\it given} bolometric
luminosity are on average a factor of $8-10$ less dust obscured at
$z\sim 2$ than at the present epoch, confirming the trend first noted
by \citet{adel00} between galaxies at $z=0$, $z\sim 1$, and $z\sim 3$;
this result is also suggested by the work of \citet{calzetti99}.
Again, this result could be anticipated if successive generations of
star formation add to already existing dust within galaxies and/or if
the dust distribution within galaxies becomes more compact with time
(e.g., via the effects of mergers which tend to drive gas and dust to
the central kpc of the system).  The net result of dust enrichment and
a more compact distribution of dust (e.g., after a merger event) is an
increase in the dust column density towards star-forming regions.  The
relationship between dust obscuration and $L_{\rm bol}$ (i.e.,
Eq~\ref{eq:bolo}) indicates that for the mean $L_{\rm bol}$ of the
$\ugr$ selected sample of galaxies of $L_{\rm bol}\sim 2.3\times
10^{11}$~L$_{\odot}$, the average dust obscuration is $\langle L_{\rm
IR}/L_{1600}\rangle \approx 4.6$.  This factor is in excellent
agreement with the mean attenuation of $4.5-5.0$ inferred from stacked
X-ray analyses \citep{reddy04}.  One would observe a factor of $4-5$
attenuation in a galaxy one order of magnitude less luminous at $z=0$
than at $z\sim 2$.  The implication is that, while it is true that a
larger fraction of star formation at high redshifts occurs in dustier
systems, the dust obscuration we observe for galaxies of a given
$L_{\rm bol}$ has less of an impact on observations of high redshift
galaxies than one would have surmised on the basis of present day
galaxies.

\section{Properties of $24$~$\mu$m Faint Galaxies}
\label{sec:undet}

\begin{figure*}[!tbp]
\plottwo{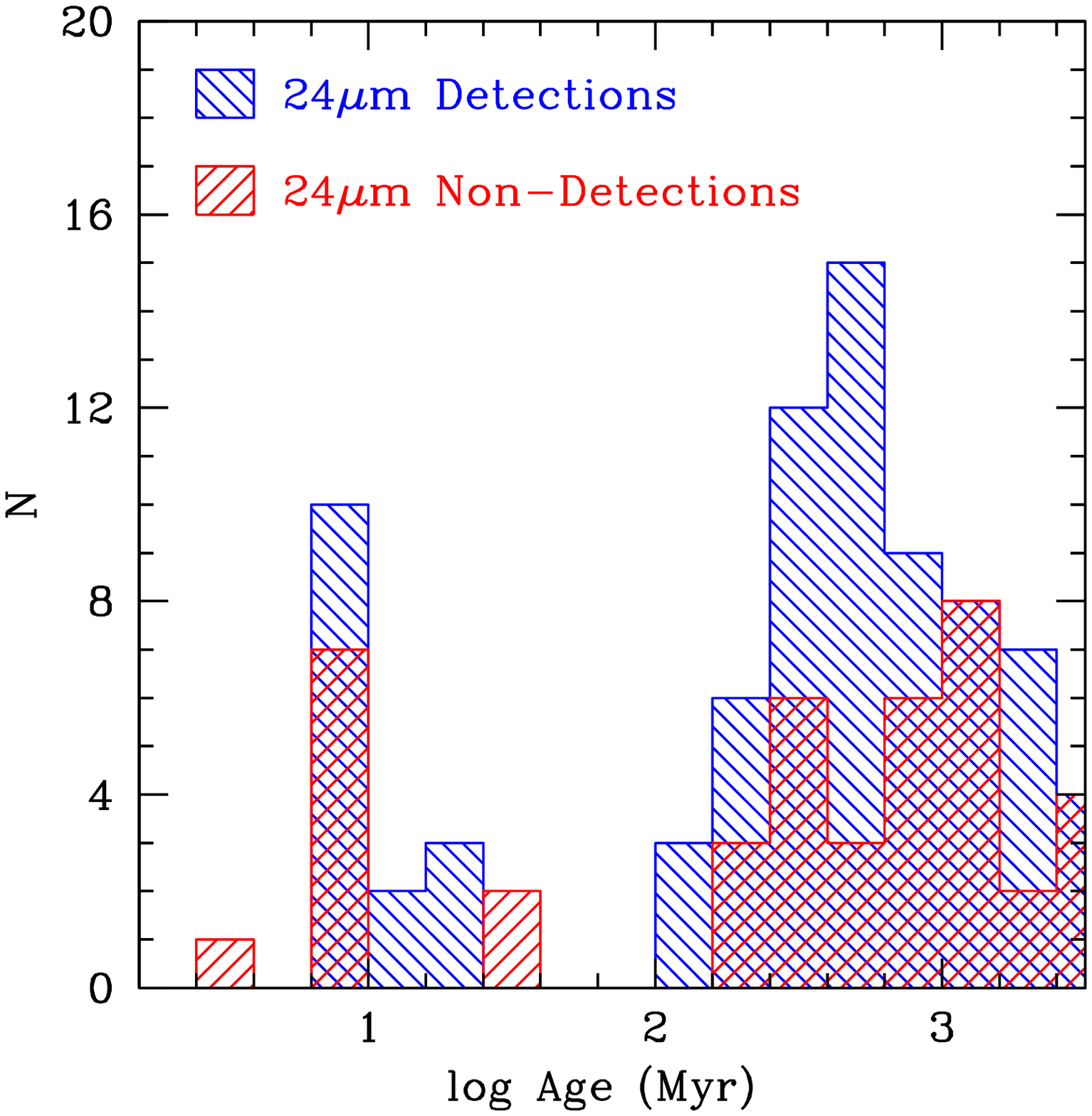}{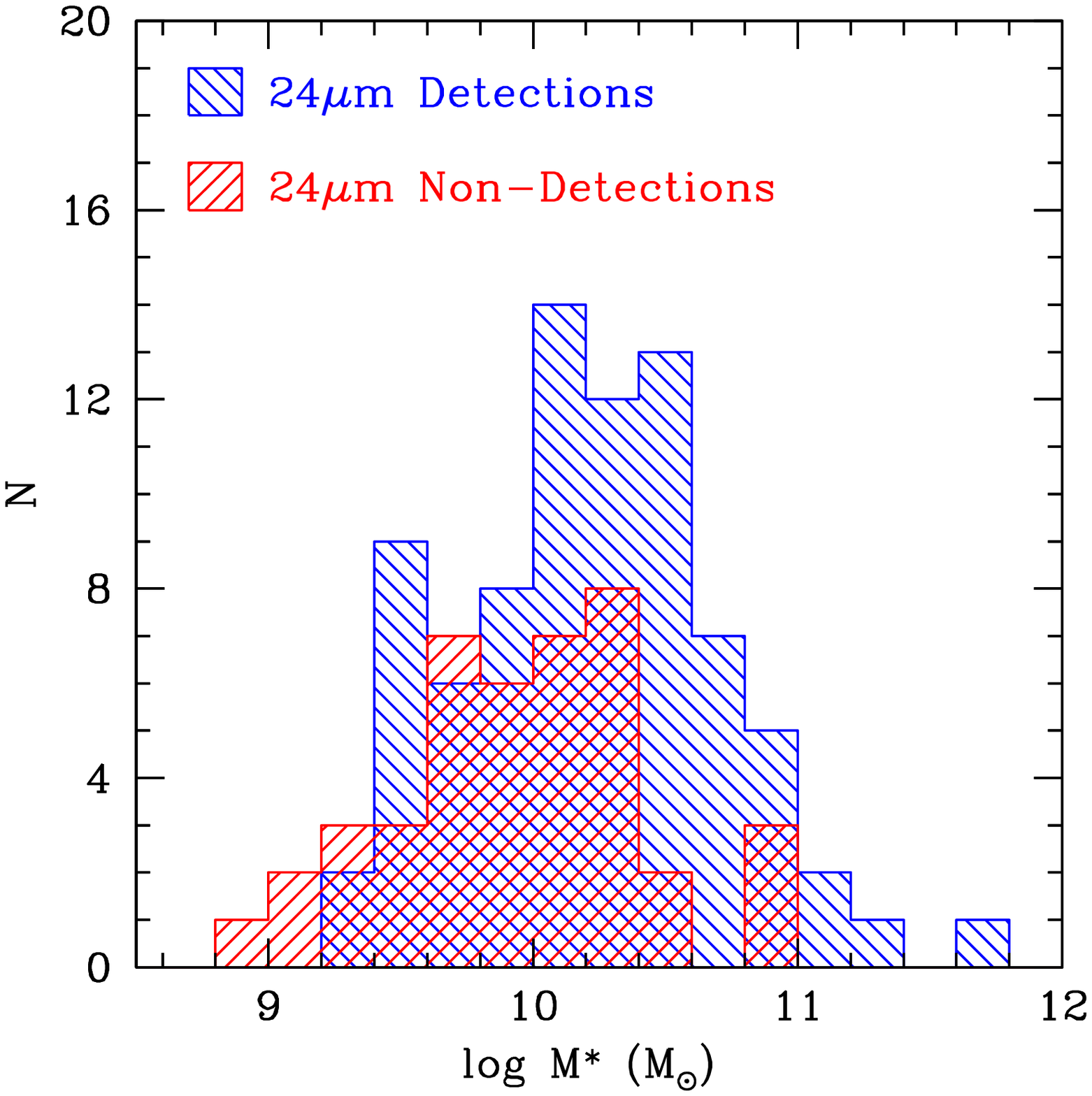}
\caption{Histograms of the age (left panel) and mass (right panel)
distributions for $24$~$\mu$m detected and non-detected $\ugr$
galaxies with redshifts $1.5<z<2.6$.  The age distributions of
detected and non-detected $\ugr$ galaxies are similar.  The mean of
the mass distributions are offset such that undetected galaxies have
$\langle \log M^{\ast}\rangle$ that is $0.4$~dex lower than that of
$24$~$\mu$m detected galaxies.\\
\label{fig:agemass}}
\end{figure*}

\subsection{Ages and Masses of Faint $24$~$\mu$m Galaxies}
\label{sec:agemassfaint}

In addition to the information gleaned from the stacking analysis
described above, we also have detailed information on the stellar
populations of galaxies with faint mid-IR emission.  Optical and
near-IR selected galaxies undetected at $24$~$\mu$m appear to have a
distribution in ages which is similar to that of $24$~$\mu$m detected
galaxies (left panel of Figure~\ref{fig:agemass}), so young ages
cannot explain why they are undetected at $24$~$\mu$m.  Alternatively,
although we find a large range in inferred stellar mass of galaxies
with $f_{\rm 24\mu m}< 8$~$\mu$Jy, the {\it mean} stellar mass of
undetected objects is $0.4$~dex lower in $\log M^{\ast}$ than $24$~$\mu$m
detected galaxies (right panel of Figure~\ref{fig:agemass}).
Regardless of these small differences in the stellar populations of
$24$~$\mu$m detected and undetected sources, the X-ray data indicate
that the primary reason why galaxies are undetected at $24$~$\mu$m is
because they have lower SFRs (Figure~\ref{fig:pahvxray}).  We
demonstrate in the next section how the rest-frame UV spectral
properties of galaxies can be used to interpret their $24$~$\mu$m
emission.

\subsection{Composite UV Spectra}
\label{sec:uvstack}

\begin{figure*}[htbp]
\includegraphics[angle=-90,width=\textwidth]{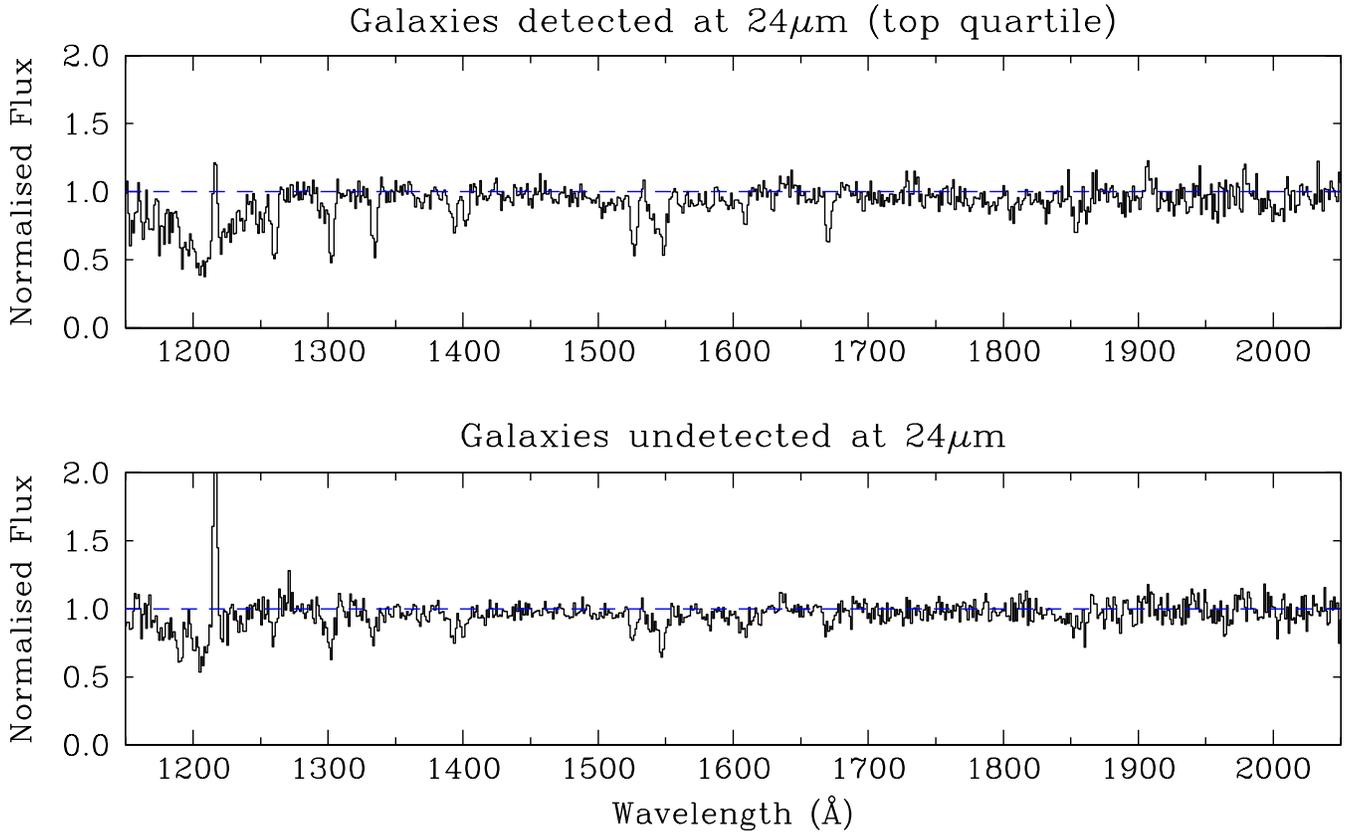}
\caption{Normalized composite UV spectra for the 39 galaxies in the
top quartile of $24$~$\mu$m detected $\ugr$ galaxies (top) with
$\langle f_{\rm 24\mu m}\rangle\sim 100$~$\mu$Jy and 73 $\ugr$
galaxies undetected at $24$~$\mu$m (bottom) with $\langle f_{\rm 24\mu
m}\rangle\sim 3$~$\mu$Jy.\\
\label{fig:comp1}}
\end{figure*}

A unique advantage of our optical $\ugr$ selection is the efficiency
with which we are able to obtain rest-frame UV spectra for these
galaxies, and this spectroscopy allows for an independent probe of the
physical conditions in the ISM.  While the S/N of any individual
spectrum is typically too low to accurately measure interstellar
absorption line widths, we can take advantage of the large number of
spectra by stacking them to create a higher S/N composite spectrum.
To investigate differences in the ISM as a function of infrared
luminosity, we constructed composite UV spectra for (a) the top
quartile of $24$~$\mu$m detected $\ugr$ galaxies, and (b) all $\ugr$
galaxies undetected at $24$~$\mu$m.  In order to stack the spectra, we
first de-redshifted them by the systemic redshift.  The systemic
redshift was inferred from a weighted combination of the measured
absorption and/or emission line redshifts, following the procedure of
\citet{adel03}.  We used the \citet{rix04} prescription to normalize
the composite spectra to the underlying stellar continua.  The
detected and undetected composite spectra consist of $39$ and $73$
galaxies, respectively, and are shown in Figure~\ref{fig:comp1}.  For
comparison, the mean $24$~$\mu$m flux of MIPS detected and undetected
galaxies is $\langle f_{\rm 24~\mu m}\rangle \sim 100$~$\mu$Jy and
$\sim 3$~$\mu$Jy, respectively; both sub-samples have $\langle
z\rangle \sim 2.1$.  Table~\ref{tab:ew} lists the measured rest-frame
equivalent widths of several interstellar absorption lines in the
composite spectra.  The primary difference between the rest-frame UV
spectra of $24$~$\mu$m detected and undetected galaxies is that the
latter have interstellar absorption lines that are a factor of 2
weaker than the lines in the $24$~$\mu$m detected galaxies.  Because
the line strengths are controlled by the combination of the velocity
spread in outflowing interstellar material and the covering fraction
of optically-thick material, this indicates that galaxies weak in
mid-IR emission are likely to have more quiescent ISM than
$24$~$\mu$m-bright galaxies, a result expected if those galaxies
undetected by MIPS have lower SFRs, and hence lower energy input into
the ISM and a lower level of dust enrichment, than $24$~$\mu$m
detected galaxies.

\begin{deluxetable}{lccc}
\tabletypesize{\footnotesize}
\tablewidth{0pc}
\tablecaption{Interstellar Absorption Line Wavelengths and Equivalent
Widths for $24$~$\mu$m Detected and Undetected $\ugr$ Galaxies}
\tablehead{
\colhead{Line} &
\colhead{$\lambda_{\rm trans}$\tablenotemark{a}} &
\colhead{Detected W$_{\lambda}$\tablenotemark{b}} &
\colhead{Undetected W$_{\lambda}$\tablenotemark{b}}}
\startdata
\\
Si II & 1260.4 & 2.58 & 0.99 \\
OI + Si II & 1303.3 & 2.27 & 1.58 \\
C II & 1334.5 & 2.05 & 0.94 \\
Si II & 1526.7 & 2.36 & 1.32 \\
Fe II & 1608.5 & 0.87 & 1.15 \\
Al II & 1670.8 & 1.85 & 1.15 \\
Al III & 1854.7 & 1.66 & 0.55 \\
Al III & 1862.8 & 0.81 & 0.64 \\
\enddata
\label{tab:ew}
\tablenotetext{a}{Transition wavelength in \AA.}
\tablenotetext{b}{Measured rest-frame equivalent width in \AA.\\}
\end{deluxetable}

Comparing the mid-IR detections with nondetections, we find the latter
have significantly stronger Ly$\alpha$ emission than the former.  The
emergent Ly$\alpha$ profiles of galaxies will depend strongly on a
number of physical parameters including the spectrum of UV radiation
(i.e., the stellar IMF), presence of outflows, and dust covering
fraction.  Neglecting all of these effects, galaxies with larger SFRs
will have stronger Ly$\alpha$ emission.  However, given that the
bolometric luminosity of star-forming galaxies scales with
dust-obscuration (e.g., Figure~\ref{fig:bolo}), we might expect
$24$~$\mu$m detected galaxies to have larger dust column densities
than undetected galaxies; this may partly explain the absence of
Ly$\alpha$ emission in $24$~$\mu$m detected galaxies.  In addition,
the velocity spread of the ISM will also affect the emergent
Ly$\alpha$ profile: the larger velocity spread in $24$~$\mu$m detected
galaxies, as indicated by their stronger interstellar absorption
lines, implies most Ly$\alpha$ photons will have larger scattering
path-lengths and are more likely to be attenuated by dust and/or
scattered out of resonance (e.g., \citealt{hansen05,adel03}).

Finally, we note that the stacked X-ray analysis of
$24$~$\mu$m-undetected galaxies confirms they have lower SFRs than
$24$~$\mu$m-detected galaxies.  Therefore, galaxies are undetected at
rest-frame $5-8.5$~$\mu$m primarily because they have lower SFRs and
not because they are deficient in mid-IR PAH luminosity for a given
$L_{\rm IR}$.  If such undetected galaxies had depressed MIR/IR flux
ratios, we would not have expected to see as large a difference in the
strengths of their interstellar absorption lines as compared with
$24$~$\mu$m detected galaxies.

\section{Mid-IR Properties of Massive Galaxies at $z\sim 2$}
\label{sec:masses}

The epoch between $z=3$ and $z=1$ appears to be the most active in
terms of the buildup of stellar mass (e.g., \citet{dickinson03}, see
also \S~\ref{sec:intro}), but significant numbers of massive galaxies
($M^{\ast}\ga10^{11}$~L$_{\odot}$) already appear to be in place by
redshifts $z\sim 2$.  The subsequent evolution of these massive
galaxies and their relation to the local population of massive and
passively evolving elliptical galaxies is an important question.  It
is useful to determine, therefore, what the mid-IR properties of
massive galaxies tell us about their bolometric luminosities.

\begin{figure}[hbt]
\plotone{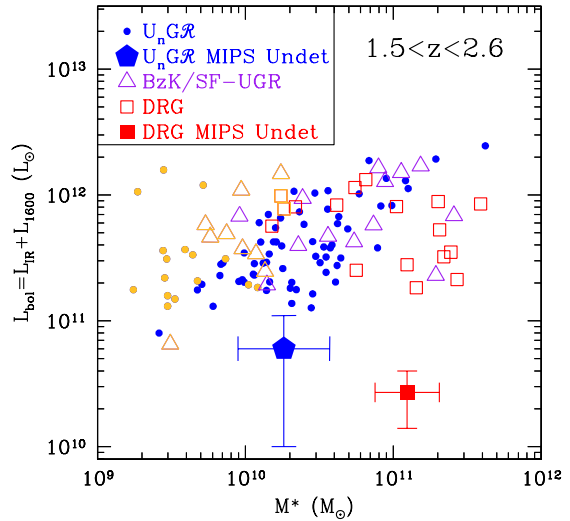}
\caption{Inferred $L_{\rm bol}$ as a function of best-fit stellar mass
assuming a CSF model for galaxies in the $\ugr$, $\bzk$, and DRG
samples.  We assumed a IR/MIR ratio of $17.2$ to convert $L_{\rm
5-8.5\mu m}$ to $L_{\rm IR}$.  The large solid pentagon and square
denote results for undetected $\ugr$ and the $7$ undetected DRG
galaxies, respectively.  The light gray (yellow) symbols indicate
$24$~$\mu$m-detected galaxies with inferred ages $\le 100$~Myr.\\
\label{fig:masses}}
\end{figure}

Figure~\ref{fig:masses} shows the bolometric luminosity of galaxies in
the $\ugr$, $\bzk$, and DRG samples as a function of best-fit stellar
mass.  The mass modeling comes from the SED analysis where we have fit
for the ${\sc}R$+$J\ks$+IRAC photometry assuming a constant star
formation (CSF) model.  As discussed elsewhere, the stellar mass is
the most well determined parameter from the SED analysis and is
relatively insensitive to the assumed star formation history
(\citealt{shapley05}; \citealt{papovich01}).  There are several
interesting aspects worth considering in Figure~\ref{fig:masses}.
First, we note that galaxies with the youngest inferred ages ($\la
100$~Myr) have low stellar masses ($M^{\ast}\la 2\times
10^{10}$~M$_{\odot}$) and span a large range in bolometric luminosity
from LIRG- to ULIRG-type galaxies.  Among $\ugr$ galaxies with
spectroscopic redshifts $1.5<z<2.6$, we note a trend of increasing
$L_{\rm bol}$ with stellar mass; $\ugr$ galaxies with $M^{\ast}\ga
10^{11}$~M$_{\odot}$ have infrared luminosities typical of ULIRGs.
The $\bzk$/SF and DRG criteria cull $M^{\ast}\ga 10^{11}$~M$_{\odot}$
galaxies with a large range in bolometric luminosity, from
$10^{11}$~L$_{\odot}$ to $\ga 10^{12}$~L$_{\odot}$.  Optically
selected sources undetected at $24$~$\mu$m have stellar masses similar
to those of detected $\ugr$ galaxies, but with $L_{\rm bol}$ a factor
of $\sim 10$ lower.  As noted before, the $24$~$\mu$m emission from these
galaxies appears to be primarily dependent on their current star
formation rate.  Finally, we note that DRGs undetected at $24$~$\mu$m
have similar stellar masses as those which are detected ($M^{\ast} \ga
10^{11}$~M$_{\odot}$), but are on average $40$ times less
bolometrically luminous.  As noted in \S~\ref{sec:lirdist}, these
$24$~$\mu$m-undetected DRGs have very red $\zmk$ colors.  Their faint
$24$~$\mu$m emission corroborates the results of X-ray stacking
analyses that indicate these galaxies have very little current star
formation \citep{reddy05a}.

The results of Figure~\ref{fig:masses} suggest that galaxies with
large stellar masses at $z\sim 2$ span a large range in bolometric
luminosity, from galaxies with little star formation to ULIRG-type
systems.  The range is likely larger than what one would infer from
Figure~\ref{fig:masses} since we have excluded directly-detected X-ray
sources that may be heavily star-forming galaxies and/or host AGN.
Figure~\ref{fig:masses} suggests that $\ga 70\%$ of massive galaxies
($M^{\ast}>10^{11}$~M$_{\odot}$) with redshifts $1.5<z<2.6$ in optical
and near-IR surveys have $L_{\rm bol}\ga 3\times 10^{11}$~L$_{\odot}$
(or SFRs $\ga 20$~$\sfr$, assuming the SFR calibration of
\citealt{kennicutt98}).  Our results imply that most $1.5<z<2.6$
galaxies with large stellar masses ($M^{\ast} \ga
10^{11}$~M$_{\odot}$) have levels of star formation that exceed those
of LIRGs.

\section{Discussion}
\label{sec:disc}

\subsection{Selection of LIRGs and ULIRGs at $z\sim 2$}

In \S~\ref{sec:lirdist} we showed that optical and near-IR selected
samples to ${\cal R}=25.5$, or $\ks=22$, host galaxies with a wide
range in infrared luminosity, from a few times $10^{10}$~L$_{\odot}$
up to the most luminous objects at these redshifts with $L_{\rm
IR}>10^{12}$~L$_{\odot}$.  Typical galaxies in these samples have
infrared luminosities in the range $10^{11}<L_{\rm
IR}<10^{12}$~L$_{\odot}$, similar in luminosity to, but with
dust-obscuration a factor of $\sim 10$ lower than (see
\S~\ref{sec:dustobs}), local LIRGs.  One advantage of optical (i.e.,
rest-frame UV) selection of LIRGs and ULIRGs at high redshift is that
it allows for the kind of efficient spectroscopic followup necessary
to accurately interpret the nature of $24$~$\mu$m emission from these
galaxies.  As shown in \S~\ref{sec:photoz}, the {\it K}-correction
depends sensitively on redshift such that even for ``good''
photometric redshift estimates with $\delta z/(1+z)\sim 0.1$, the
corresponding uncertainty in the rest-frame $5-8.5$~$\mu$m luminosity
increases by a factor of 5.  A particularly unique aspect of our study
combining the optical sample with MIPS observations is that the
spectroscopic database can be used to assess the physical conditions
of the ISM in galaxies as function of $L_{\rm bol}$, providing an
additional method for probing the detailed nature of $24$~$\mu$m
galaxies at $z\sim 2$ (\S~\ref{sec:uvstack}).  Aside from the
constraints on the mid-IR luminosities possible with spectroscopic
redshifts, precise positions of sources from higher spatial resolution
and shorter wavelength data enable the deblending of most $z\sim 2$
galaxies.  The deblending procedure made possible by optical, near-IR,
and {\it Spitzer} IRAC observations enable accurate identification and
photometry of faint galaxies well below the MIPS $24$~$\mu$m confusion
limit and will provide a more complete ``census'' of the LIRG
population at redshift $z\sim 2$ than possible using MIPS observations
alone.

Further, selection by optical colors gives important information on
the {\it unobscured} component of the star formation in galaxies and
complements well the information on the obscured component probed by
the $24$~$\mu$m observations.  Objects with lower star formation rates
will have bolometric luminosities that are typically dominated by the
observed UV emission and objects with larger star formation rates will
have bolometric luminosities that are dominated by the observed IR
emission.  The transition between the UV and IR dominated regimes
(i.e., where $L_{\rm IR}=L_{1600}$) at $z\sim 2$ occurs for galaxies
with $L_{\rm IR}\approx 10^{11}$~L$_{\odot}$, or about $0.3L^*$
(Figure~\ref{fig:bolo}).  A comparison with the $z=0$ sample of
\citet{bell03} shows that the bolometric luminosities begin to be
dominated by IR emission (i.e., $L_{\rm IR}/L_{1600}>1$) for galaxies
which are two orders of magnitude more luminous at $z\sim 2$ than at
the present epoch.  As discussed in \S~\ref{sec:dustobs}, this is
plausibly explained as a result of higher dust-to-gas ratios in the
local galaxies.  More generally, galaxies of a given dust obscuration
are anywhere from 2-100 times more luminous at $z\sim 2$ than locally
(Figure~\ref{fig:bolo}), with the greatest difference for galaxies
with relatively low $L_{\rm IR}/L_{1600}\la 20$.  The implication of
these observations is that while it is certainly true that a larger
fraction of the star formation at high redshifts occurs in dustier
galaxies, selection via rest-frame UV colors (and performing followup
spectroscopy) is easier at high redshift than locally for galaxies at
a given bolometric luminosity.  Optical selection is therefore
arguably the most promising and spectroscopically efficient method for
selecting LIRGs (which undoubtedly account for a significant fraction
of the star formation rate density and far-infrared background; e.g.,
\citealt{adel00}) at $z\sim 2$.

As demonstrated in \S~\ref{sec:lirdist} and in \citet{reddy05a},
ULIRGs and SMGs at these redshifts also often appear in optical and
near-IR selected samples; $\sim 50\%$ of the most luminous SMGs have
enough unobscured star formation that they satisfy the $z\sim 2$
optical criteria \citep{chapman05}.  Accounting for the obscured
portion of star formation in these ultraluminous sources of course
requires the longer wavelength data since rest-frame UV slopes
underpredict their $L_{\rm IR}$ (\S~\ref{sec:attenuation}).  In some
sense, $24$~$\mu$m observations are a more powerful method of
estimating $L_{\rm bol}$ of these ultra-luminous sources since their
detection significance at $24$~$\mu$m is typically $50-100$ times
larger and the beamsize is a factor of $9$ smaller than at
$850$~$\mu$m.  The ability to uniformly cover large areas of the sky
with MIPS observations provides an additional advantage over current
submillimeter surveys which suffer from areal and depth
incompleteness, thus making it difficult to constrain the volumes
probed.  Using $24$~$\mu$m observations to assess the global
energetics of ultraluminous sources of course requires that we
accurately calibrate the $L_{\rm 5-8.5\mu m}/L_{\rm IR}$ ratios for
these objects.

We have demonstrated that the typical galaxy in optical and near-IR
samples of $z\sim 2$ galaxies has $L_{\rm IR}$ corresponding to that
of LIRGs.  A related issue is whether most LIRGs at $z\sim 2$ can be
selected by their rest-frame UV or optical colors.  A direct
comparison of the number counts of MIPS sources to the number of
$24$~$\mu$m detected $\ugr$ and $\bzk$ galaxies to $8$~$\mu$Jy (the
GOODS-N MIPS $24$~$\mu$m $3$~$\sigma$ sensitivity limit) is not
possible since (a) we primarily relied on the $\ks$-band data to
deblend sources in the $24$~$\mu$m imaging and (b) the redshift
distribution of MIPS sources to $8$~$\mu$Jy is not yet well
established.  Nonetheless, including both optical and near-IR selected
LIRGs at $z\sim 2$ ensures that we must be reasonably ``complete'' for
both optically-bright (and $24$~$\mu$m-faint) LIRGs in the optical
sample and optically-faint (and $24$~$\mu$m-bright) LIRGs in the
near-IR selected $\bzk$ and DRG samples.  Galaxies with LIRG
luminosities will predominantly have near-IR magnitudes bright enough
to be considered in our analysis (i.e., $\ks\la 21$ according to the
left panel of Figure~\ref{fig:lircol}).  Objects not selected by these
various criteria will likely either fall at different redshifts, not
have LIRG luminosities, and/or may be scattered out of the color
selection windows due to photometric error (e.g., \citealt{reddy05a}).
As an example of one form of photometric scatter, in the course of the
$z\sim 3$ Lyman Break Galaxy Survey, we relied on $\ugr$ photometry
based on images of the HDF-North field taken at the Palomar Hale $5$~m
Telescope \citep{steidel03}.  We subsequently imaged a larger portion
of the GOODS-North field using the Keck I Telescope \citep{steidel04}.
Our photometric analysis indicates that of the BX/BM objects to ${\cal
R}=25.5$ identified in the Palomar imaging, about $76\%$ were
recovered as BX/BM objects in the Keck imaging.  A small fraction of
the remaining $24\%$ were recovered using LBG selection.  The level of
scatter between different photometric realizations in other fields is
also typically $\sim 25\%$ and is mostly due to the narrow photometric
windows used to select BX/BM galaxies.  Regardless of these
photometric effects, it is highly unlikely that there exist large
numbers of LIRGs at $z\sim 2$ with such different optical and near-IR
properties that they would be completely absent from all the samples
considered here.  Finally, our knowledge of the exact positions of
optical and near-IR selected galaxies from the higher spatial
resolution $\ks$-band and IRAC data allows us to mitigate the effects
of confusion (see \S~\ref{sec:midir}), so we should be reasonably
complete for galaxies which are detected at $24$~$\mu$m to $8$~$\mu$Jy
but which might otherwise be confused with brighter sources.  It is
therefore reasonable to conclude that the LIRG population at $z\sim 2$
is essentially the same population of galaxies that are selected in
optical and near-IR samples.

\subsection{Mass Assembly at High Redshift}

\begin{figure*}[hbt]
\plotone{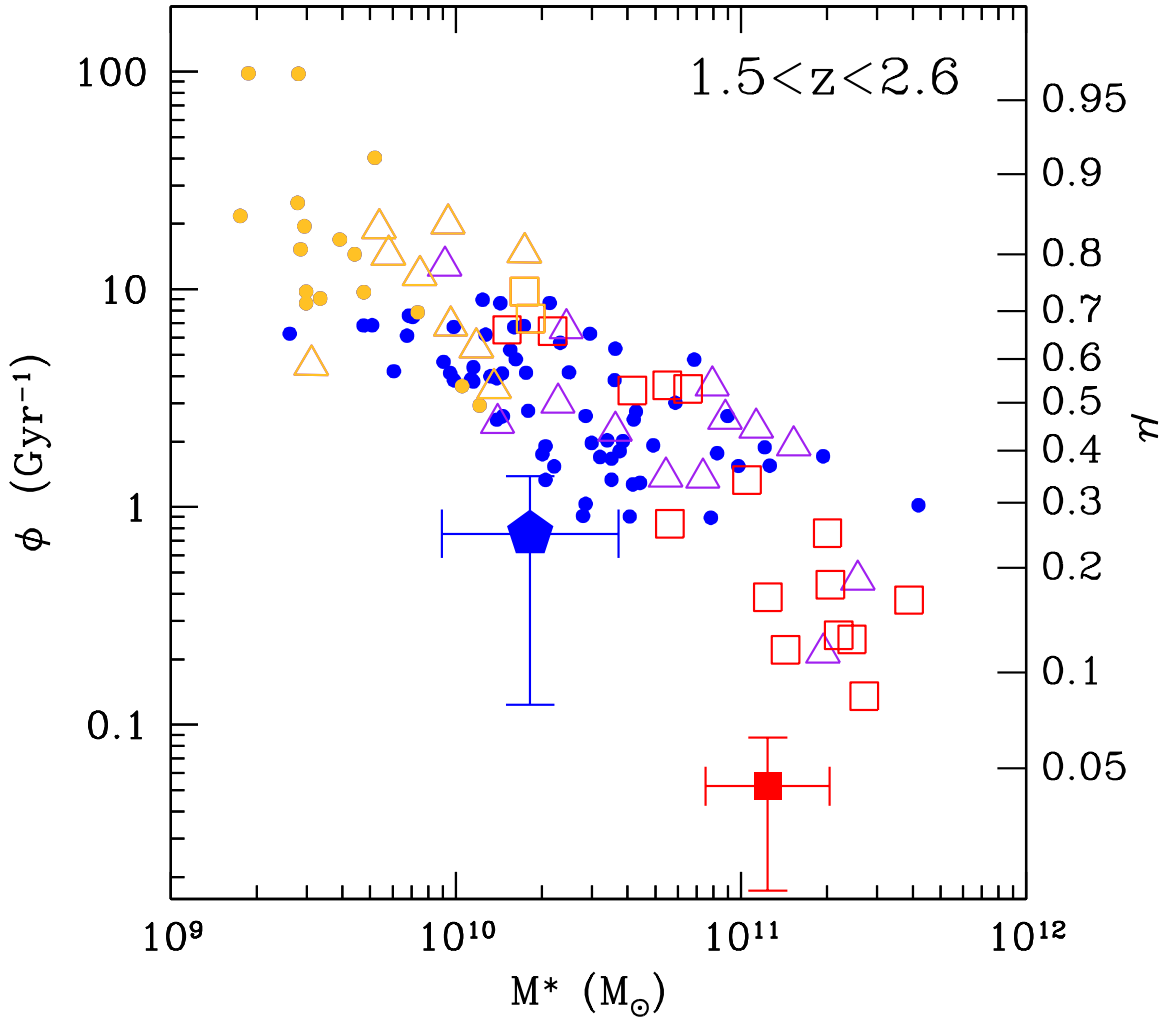}
\caption{Specific SFR, $\phi$, as a function of stellar mass for
galaxies in the samples considered here.  Symbols are the same as in
Figure~\ref{fig:masses} (yellow symbols indicate galaxies with
inferred ages $\la 100$~Myr).  The right-hand axis indicates the gas
fraction $\mu$ associated with galaxies of a given $\phi$.\\
\label{fig:specsfr}}
\end{figure*}

We demonstrated that LIRGs and ULIRGs are present over the full range
of stellar mass, from $\sim 2\times10^{9}$~M$_{\odot}$ to
$5\times10^{11}$~M$_{\odot}$, for galaxies in the samples considered
here (Figure~\ref{fig:masses}).  To assess the significance of the
current star formation in the buildup of stellar mass, we have
computed the specific star formation rate, $\phi$, defined as the SFR
per unit stellar mass.  We show the observed $\phi$ for galaxies in
our sample as function of stellar mass in Figure~\ref{fig:specsfr}.
The correlation between $\phi$ and $M^{\ast}$ could have been
predicted from Figure~\ref{fig:masses} since the range of $L_{\rm
bol}$ is similar over the range of $M^{\ast}$ considered here.  We
also note that the correlation is accentuated since (a) $\phi$ is not
independent of $M^{\ast}$ and (b) there are presumably galaxies with
low $\phi$ and low $M^{\ast}$ that would be missing from the optical
and near-IR samples (irrespective of the MIPS detection limit).
Furthermore, the upper envelope of points in Figure~\ref{fig:specsfr}
is defined by our cut to exclude luminous AGN based on the {\it
Chandra} X-ray data.  Nonetheless, we find that star-forming galaxies
with large stellar masses $M^{\ast}\ga 10^{11}$~M$_{\odot}$ without an
AGN signature have specific SFRs that are $1$ to $2$ orders of
magnitude lower than those of young galaxies (yellow symbols),
implying that the current star formation contributes more
significantly to the buildup of stellar mass in low mass galaxies than
high mass galaxies at $z\sim 2$.  This change in specific SFR as a
function of mass has been observed at later epochs as well (e.g.,
\citealt{bell05b}).  The shifting of the relationship to lower
specific SFRs at later epochs for galaxies with large stellar masses
has been referred to as ``downsizing'' \citep{cowie96}.

The most massive, star-forming DRGs at these redshifts with large
dust-to-gas ratios (see Figure~\ref{fig:bolo}) also have the lowest
$\phi$ for MIPS-detected galaxies at $z\sim 2$.  \citet{papovich05}
demonstrate that the integrated specific SFR of these massive DRGs at
$z\sim 2$ is $\sim 2$ magnitudes larger than for galaxies with similar
masses ($M^{\ast}>10^{11}$~M$_{\odot}$) at lower redshifts $0.3\le z<0.5$,
based on a comparison with the COMBO-17 sample \citep{bell05b}.  This
decrease in $\phi$ suggests that $M^{\ast}>10^{11}$~M$_{\odot}$ galaxies
have built up most of their stellar mass by $z\sim 2$.  In fact, the
mass-doubling time becomes increasingly large for these massive galaxies
if we assume the case of an exponentially declining star formation
model.  In this case, $\phi$ will evolve with time $t$ as
\begin{eqnarray}
\phi(t) = {{(1+f)\exp(-t/\tau)}\over{\tau[1-\exp(-t/\tau)]}},
\label{eq:specsfr}
\end{eqnarray}
where $f$ is the ratio of the gas mass that is lost due to outflows to
that formed in stars and $\tau$ is the star formation decay timescale
as defined in \S~\ref{sec:photoz}.  Given some initial SFR, a CSF
model (i.e., with $\tau=\infty$) will track straight through the
points corresponding to galaxies with $M^{\ast}\ga
10^{11}$~M$_{\odot}$, but such a model would predict a $\phi$ at
$z\sim 0.4$ that is $\sim 5$ times larger than observed at $z\sim 0.4$
in the COMBO-17 survey \citep{bell05b}.  Therefore, a declining
star-formation history may be more appropriate for describing the
future evolution of galaxies with large stellar masses at $z\sim 2$.
In particular, \citet{erb06a} find that a model that assumes a
super-solar yield of metals (i.e., ratio of mass of metals ejected
into the ISM to mass of metals locked in long-lived stars) of
$y=1.5$~Z$_{\odot}$ and an outflow rate of $4\,\times\,{\rm SFR}$
($f=4$) appears to best-fit the observed metallicities of $\ugr$
galaxies at $z\sim 2$ as a function of gas fraction.  If all
star-forming galaxies at $z\sim 2$ follow a similar evolutionary track
as the $M^{\ast}>10^{11}$~M$_{\odot}$ galaxies (i.e., follow an
exponentially declining star formation history with large outflow
rate), then the scatter of galaxies with a given specific SFR simply
reflects the range in the final stellar masses and dark matter halo
masses (see also discussion in \citealt{erb06a}).

We can directly relate the specific SFR $\phi$ with the cold gas
fraction, $\mu \equiv M_{\rm gas}/(M_{\rm
gas}+M^{\ast})$.
If we assume that the SFR is proportional to the cold gas mass,
$M_{\rm gas}$, to the $1.4$ power according to the Schmidt law
\citep{kennicutt98b}, then
\begin{eqnarray}
\phi \propto {{M^{1.4}_{\rm gas}}\over{M^{\ast}}}.
\end{eqnarray}
It then follows that 
\begin{eqnarray}
\phi = C{{\mu^{1.4}}\over{1-\mu}},
\end{eqnarray}
where $C$ is a constant that depends on the constant of
proportionality between the SFR and gas mass surface densities in the
Schmidt law and the total gas mass at virialization (i.e., when star
formation commences).  There is a one-to-one correspondence between
the specific SFR, $\phi$, and gas fraction, $\mu$, such that galaxies
with large specific SFRs will have a larger fraction of cold gas than
galaxies with small specific SFRs.  \citet{erb06b} demonstrate that
$\mu$ decreases as a function of stellar mass for a large sample of
$\ugr$ galaxies with H$\alpha$ spectroscopy, with a mean $\mu$ across
the sample of $\langle\mu\rangle\sim 0.5$.  The mean specific SFR for
these galaxies is $\langle\phi\rangle\sim 3$~Gyr$^{-1}$.  Using these
mean values to estimate $C$, we show the range of $\mu$ on the
right-hand axis of Figure~\ref{fig:specsfr}.  Young galaxies with ages
less than $100$~Myr in our sample also have the largest gas fractions
($\mu\sim 0.6-0.9$) and largest specific SFRs
(Figure~\ref{fig:specsfr}) compared with older galaxies at $z\sim 2$.
These results strongly suggest that the young galaxies have large
reservoirs of cold gas and have just begun forming stars.  Similarly,
galaxies with the largest stellar masses have lower specific SFRs and
therefore lower cold gas fractions ($\mu\sim 0.1-0.3$) and are likely
to cease star-formation in a relatively short time.  The correlation
between $\phi$ and $M^{\ast}$ revealed by Figure~\ref{fig:specsfr} is
actually expected from the trend between $\mu$ and $M^{\ast}$ inferred
from the H$\alpha$ spectroscopic analysis of \citet{erb06b}.
Furthermore, \citet{erb06a} find a tight trend between metallicity and
gas fraction such that galaxies with lower $\mu$ are more metal-rich.
These galaxies will therefore be more dust-obscured since metallicity
is directly proportional to dust-to-gas ratio.  Our analysis confirms
our expectation that the most dust-obscured objects at $z\sim 2$
(i.e., the DRGs with the largest dust-to-gas ratios,
Figure~\ref{fig:bolo}) also have low gas fractions as demonstrated in
Figure~\ref{fig:specsfr}.  Figure~\ref{fig:specsfr} indicates that the
trend in gas fraction versus stellar mass (or, metallicity versus gas
fraction) found for $\ugr$-selected galaxies with H$\alpha$
spectroscopy \citep{erb06a} also applies to galaxies selected by their
near-IR colors.

Combining the optical and near-IR samples, we find a wide range in the
evolutionary state of galaxies at $z\sim 2$, from those which are just
starting to form stars to those which have already accumulated most of
their stellar mass and are about to become passive or already are.
Almost all DRGs with the reddest $\zmk$ color ($\zmk>3$) are
undetected at $24$~$\mu$m and in the deep {\it Chandra} X-ray data
(e.g., see right panel of Figure~\ref{fig:lircol} and Figure~16 in
\citealt{reddy05a}).  The specific SFRs of these red DRGs are the
lowest observed for the $z\sim 2$ galaxies considered here; they have
cold gas fractions of less than $5\%$, confirming that they have
essentially shut-off star formation by $z\sim 2$.  The star-forming
DRGs with $M^{\ast}>10^{11}$~M$_{\odot}$ (which are dusty and
metal-rich as inferred from their large dust-to-gas ratios; cf.,
Figure~\ref{fig:bolo}) are likely to reach this passively-evolving
state by redshifts $z\la 1.5$.  A simple model which assumes a high
outflow rate proportional to the SFR and exponentially declining star
formation history (i.e., the model which appears to best-fit the
observed metallicities of galaxies as function of gas fraction;
\citealt{erb06a}) is sufficient to explain the presence of galaxies
with large stellar masses and little star formation by $z\sim 2$
without invoking AGN feedback, despite the fact that a large AGN
fraction of $\sim 25\%$ is observed among galaxies with large stellar
masses \citep{reddy05a}.

\section{Conclusions}
\label{sec:conc}

We use {\it Spitzer} MIPS data to examine the bolometric luminosities
and extinction properties of optical and near-IR selected galaxies at
redshifts $1.5\la z\la 2.6$ in the GOODS-North field.  At these
redshifts, the mid-IR ($5-8.5$~$\mu$m) features associated with
polycyclic aromatic hydrocarbon (PAH) emission, which are ubiquitous
in local and $z\sim 1$ star-forming galaxies, are shifted into the
MIPS $24$~$\mu$m filter.  Extensive multi-wavelength data in the
GOODS-North field, including very deep {\it Chandra} X-ray data, allow
us to test the validity of inferring the bolometric luminosities of
high redshift galaxies from their rest-frame $5-8.5$~$\mu$m emission.
Galaxies at $z\sim 2$ are selected by their optical ($\ugr$) and
near-IR ($\bzk$ and $\jmk$) colors, and for comparison we also
consider radio-detected submillimeter galaxies (SMGs;
\citealt{chapman05}).  The optically-selected sample is advantageous
because we have knowledge of the precise redshifts of $\ga 300$
galaxies from spectroscopy, allowing for the most accurate estimates
of the rest-frame mid-IR luminosities of $z\sim 2$ galaxies.  In
addition to our spectroscopic sample, we use deep optical, near-IR,
and {\it Spitzer} IRAC data to derive photometric redshifts for $\bzk$
galaxies and those with red $\jmk>2.3$ colors (Distant Red Galaxies;
DRGs).  The principle conclusions of this study are as follows:

1. Using local templates to ${\it K}$-correct the observed $24$~$\mu$m
fluxes, we find that the rest-frame $5-8.5$~$\mu$m luminosity ($L_{\rm
  5-8.5\mu m}$) of $z\sim 2$ galaxies correlates well with their
stacked X-ray emission.  A subset of galaxies with H$\alpha$
measurements have H$\alpha$-inferred bolometric luminosities that
correlate very well with their $L_{\rm 5-8.5\mu m}$-inferred
bolometric luminosities (with a scatter of $0.2$~dex).  These
observations suggest that $L_{\rm 5-8.5\mu m}$ provides a reliable
estimate of $L_{\rm IR}$ for most star forming galaxies at $z\sim 2$.

2.  We find that the optical and near-IR selected $z\sim 2$ galaxies
span a very wide range in infrared luminosity from LIRG to ULIRG
objects.  We find a mean infrared luminosity of $\langle L_{\rm
  IR}\rangle \sim 2\times 10^{11}$~L$_{\odot}$ for galaxies in the
optical and near-IR samples, in excellent agreement with the value
obtained from a stacked X-ray analysis.  The optical and near-IR
selected star-forming galaxies likely account for a significant
fraction of the LIRG population at $z\sim 2$.  Galaxies
with $\ks<20$ have $L_{\rm IR}$ greater than $2$ times that of
galaxies with $\ks>20.5$.  Non-AGN galaxies with the reddest near-IR
colors ($\zmk>3$) are mostly undetected at $24$~$\mu$m suggesting they
have low SFRs, a conclusion supported by stacked X-ray analyses.

3. We demonstrate using $24$~$\mu$m and X-ray stacking analyses that
galaxies undetected to $f_{\rm 24\mu m}\sim 8$~$\mu$Jy are faint
because they have lower SFRs and/or lower obscuration than $24$~$\mu$m
detected galaxies, and not because they are deficient in PAH emission
at a given $L_{\rm IR}$.  We infer that typically half of the
bolometric luminosity of these $24$~$\mu$m-undetected galaxies is
emitted in the UV.  Comparing the rest-frame UV composite spectra of
$24$~$\mu$m-undetected galaxies with those in the top quartile of
detected objects shows that the latter have low-ionization
interstellar absorption lines that are $\sim 2$ times stronger than
the former, indicating some combination of more turbulent interstellar
media and a larger neutral gas covering fraction.  This is consistent
with our conclusion that galaxies undetected at $24$~$\mu$m have lower
SFRs than $24$~$\mu$m-detected galaxies, and therefore a lower input
of kinetic energy and dust into their ISM.

4. The $24$~$\mu$m (and deep X-ray) data indicate that galaxies whose
current star-formation episodes are older than $100$~Myr and have
infrared luminosities $10^{10}\la L_{\rm IR}\la 10^{12}$~L$_{\odot}$
appear to follow the local relation between rest-frame UV slope and
dust obscuration, implying that such galaxies at $z\sim 2$ have
moderate amounts of dust extinction and that their UV slopes can be
used to infer their extinction.  Galaxies younger than $100$~Myr have
rest-frame UV colors that are redder than expected given their
inferred $L_{\rm IR}$, indicating they may obey a steeper extinction
law.  These young galaxies have the lowest stellar masses, but span
the same range in bolometric luminosity as galaxies with larger
stellar masses.

5. Galaxies with $L_{\rm bol}\ga 10^{12}$~L$_{\odot}$, including
radio-detected SMGs, are typically $\sim 10-100$ times more dust
obscured than their UV spectral slopes would indicate, assuming their
$850$~$\mu$m-inferred infrared luminosities.  The $24$~$\mu$m-inferred
infrared luminosities of radio-detected SMGs are systematically a
factor of $2-10$ times lower than those predicted by their
$850$~$\mu$m fluxes; adopting the $24$~$\mu$m estimates implies dust
attenuation factors that are $\sim 5-50$ times larger than their UV
spectral slopes would indicate.  Regardless, such galaxies will often
be blue enough to satisfy the $\ugr$ criteria, so finding these dust
obscured galaxies in optical surveys is not uncommon.

6. A comparison between the dust obscuration in $z\sim 2$ and $z=0$
galaxies suggests that galaxies of a {\it given} bolometric luminosity
are much less dust obscured (by a factor of $\sim 8-10$) at high
redshift than at the present epoch.  This result is expected (a) as
galaxies age and go through successive generations of star formation
and dust production and (b) if the distribution of dust and star
formation in galaxies becomes more compact over time (e.g., through
mergers or interactions) resulting in greater dust column densities
towards star-forming regions.  We find that star-forming Distant Red
Galaxies (DRGs) with stellar masses $M^{\ast} \ga 10^{11}$~M$_{\odot}$ and
ages $\ga 2$~Gyr have bolometric luminosities and dust-obscuration
factors similar to those of local ULIRGs, suggesting that such DRGs,
like local ULIRGs, carry relatively large amounts of dust into their
current episodes of star formation.

7. Galaxies with the largest stellar masses at $z\sim 2$ ($M^{\ast}\ga
10^{11}$~M$_{\odot}$) also span a large range in bolometric
luminosity, from those which have red near-IR colors ($\zmk>3$) with
little current star formation to those ULIRG objects found among
optical and near-IR selected massive galaxies.  Our results suggest
$\ga 70\%$ of massive galaxies ($M^{\ast}\ga 10^{11}$~M$_{\odot}$) in
optical and near-IR surveys with redshifts $1.5<z<2.6$ have $L_{\rm
  bol}\ga 3\times 10^{11}$~L$_{\odot}$ (SFRs $\ga 20$~$\sfr$),
comparable to and exceeding the luminosity of LIRGs.  

8. Similar to lower redshift studies, we find a trend between specific
SFR (SFR per unit stellar mass) and stellar mass at $z\sim 2$ which
indicates that the observed star formation contributes more to the
buildup of stellar mass in galaxies with low stellar masses than in
those with larger stellar masses at $z\sim 2$.  This trend between
specific SFR and stellar mass indicates a strong decrease in cold gas
fraction as function of stellar mass, consistent with results from
near-IR spectroscopic observations, and suggests that galaxies with
large stellar masses ($M^{\ast}>10^{11}$~M$_{\odot}$) at $z\sim 2$
will quickly cease star formation.  Combining optical and near-IR
selected samples, we find a large range in the evolutionary state of
galaxies at $z\sim 2$, from those which have just begun to form stars
and which have large gas fractions to those which are old, massive,
and have little remaining cold gas.

\acknowledgements

We thank Natascha F${\rm \ddot{o}}$rster Schreiber and Helene Roussel
for providing their ISO spectra of local galaxies, and David Elbaz for
providing an electronic catalog of the mid-IR and IR luminosities of
local star-forming galaxies.  We also thank the referee, Emeric Le
Floc'h, for useful comments.  The work presented here has been
supported by grant AST 03-07263 from the National Science Foundation
and by the David and Lucile Packard Foundation.  We made use of the
NASA/IPAC Extragalactic Database (NED), which is operated by the Jet
Propulsion Laboratory, California Institute of Technology, under
contract with NASA.

\bibliographystyle{apj}

\end{document}